\documentstyle[12pt,epsfig]{article}
\newcommand{\M}{{\cal M}}
\newcommand{\MeV}{{\rm MeV}}
\newcommand{\GeV}{{\rm GeV}}

\newcommand{\fm}{{\rm fm}}
\renewcommand{\Im}{{\rm Im}}
\renewcommand{\Re}{{\rm Re}}

\newcommand{\lsim}{$\raisebox{-0.8ex} {$\stackrel{\textstyle <}{\sim}$}$}

\begin{document}
\title{Current correlation functions, QCD sum rules \\
and vector mesons in baryonic matter \thanks{Work supported in part by GSI and BMBF}}
\author{F. Klingl, N. Kaiser and W. Weise\\ Physik-Department\\ Technische Universit\"at
M\"unchen\\ Institut f\"ur Theoretische Physik\\ D-85747 Garching, Germany}
\date{}
\maketitle
\begin{abstract}
  Based on an effective Lagrangian which combines chiral SU(3) dynamics with
  vector meson dominance, we have developed a model for the forward vector
  meson-nucleon scattering amplitudes. We use this as an input to calculate the
  low energy part of the current-current correlation function in nuclear
  matter.  Its spectrum enters directly in the ``left hand side'' of QCD sum
  rules. For the isovector channel we find a significant enhancement of the
  in-medium spectral density below the $\rho$ resonance while the $\rho$ meson
  mass itself changes only slightly. The situation is different in the
  isoscalar channel, where the mass and peak position of the $\omega$ meson
  move downward while its width increases less drastically than in the $\rho$
  meson case. For the $\phi$ meson we find almost no mass shift; the width of
  the peak broadens moderately. We observe a remarkable degree of consistency
  with the operator product expansion of QCD sum rules in all three channels.
  We point out, however, that these results cannot simply be interpreted, as
  commonly done, in terms of a universal rescaling of vector meson masses in
  matter.
\end{abstract}

\section{\label{1} Introduction}
At present there is a lively discussion about the behaviour of vector mesons in
dense and hot hadronic matter. According to Brown and Rho \cite{1}, vector
meson masses $m_V$ (in fact the masses of most hadrons with the exception of
the pion) should drop in matter according to a simple (BR) scaling law:
$m_V^*/m_V=f_\pi^*/f_\pi$, where $f_\pi$ is the pion decay constant and the
asterix refers to in-medium quantities. A decreasing $f_\pi^*$ (more precisely;
the $f_\pi^*$ related to the time component of the axial current) in matter
indicates a tendency toward chiral restoration (the transition from the
spontaneously broken Nambu-Goldstone to the unbroken Wigner-Weyl realization of
chiral symmetry). Dropping vector meson masses would then be a signal for this
transition, if BR-scaling holds. Other scaling laws \cite{1a} using a bag model
find a mass reduction of two third times that of the nucleon mass which leads
to a similar shift as BR-scaling.

Several analyses of the $\rho$ and $\omega$ meson masses in matter using QCD
sum rules \cite{2,3} seem to confirm these expectations. On the other hand, model
calculations of the density dependent two-pion self-energy of the rho meson in
nuclear matter \cite{4,5} suggest only marginal changes of the in-medium rho
meson mass, but a strongly increased decay width instead. Additional BR scaling
seems to be needed in order to match the hadronic models with the QCD sum rule
analysis \cite{6}. Phenomenologically, a dropping $\rho$ meson mass seems to help \cite{7,8}
understanding the enhanced dilepton yields seen at masses below the rho meson
resonance in the CERES \cite{9} and HELIOS \cite{10} experiments at CERN.

Few attempts have been made so far to perform a realistic calculation of the
omega meson self-energy in matter. The influence of the density dependent
effective nucleon mass in the particle-hole polarization coupled to the
$\omega$ meson has been studied in ref.\cite{10a}. However, leading in-medium
effects from the $\omega$-decay into the three-pion continuum in the presence
of nucleons have not yet been investigated. One of the aims of this paper is to
take such effects into account.

The $\phi$ meson as the third neutral member of the SU(3) vector meson octet is special in the sense that it is
almost a pure strange quark-antiquark system. The large strange quark
mass protects it from effects related to chiral restoration. Therefore the
$\phi$ meson mass does not
seem to be rescaled in matter as strongly as the $\rho$ or $\omega$ meson when
analyzed using in-medium QCD sum rules.  

The QCD sum rule approach, at any practical level so far, has used only a
caricature of the true spectral functions, namely a parameterization in terms
of a delta function at the vector meson pole accompanied by a step function
continuum at higher energies. While such a parameterization turns out to be
valid in vacuum (as we shall confirm), this is shown not to be justified for
the in-medium spectra. In particular, the parameterization of the rho meson in
matter as an isolated quasiparticle pole becomes unreliable with increasing
baryon density, given the strong increase of its partial decay width into $\rho
N \to \pi N,\, \pi\pi N$ etc. channels.

The purpose of this paper is to present a systematic calculation of
current-current correlation functions in baryonic matter, based on an effective
Lagrangian which combines chiral symmetry with vector meson dominance of the
electromagnetic current. In the isovector channel we find a significant
enhancement of the low mass part of the spectrum with increasing baryon
density. On the other hand the $\rho$ meson mass changes at a much slower rate
with density than naively expected. In the isoscalar channel the situation
turns out to be different. The position of the $\omega$ meson pole moves
downward with rising density. The $\omega$ width does not increase as strongly
as for the isovector channel. To leading order in the density, these features
can be translated into statements about the (complex) effective scattering
lengths of $\rho$ and $\omega$ interacting with a nucleon. These will be
discussed in detail. The $\phi$ meson-nucleon scattering amplitude is much smaller
than the $\rho N$ and $\omega N$ amplitudes, with a slight increase of the
width but almost no shift of the resonance. We show that the vector
meson-nucleon scattering amplitudes derived from our hadronic effective
Lagrangian lead to spectral densities which are remarkably consistent with the
standard QCD sum rule analysis, without the need for additional BR scaling.

This paper is organized as follows. In section \ref{2} we review properties of
the current correlation functions in vacuum and give a brief account of the QCD
sum rule approach. The current-current correlation function in baryonic matter
will be developed in section \ref{3}. To leading order in the density this
requires a detailed calculation of the vector meson-nucleon scattering
amplitudes which will be described in section \ref{4}. The in-medium spectra
and their comparison with QCD sum rules is presented in section \ref{5}, and
the results will be discussed in section \ref{6}.

\section{\label{2} Current-current correlation functions in vacuum}
\subsection{Basics}
This is an introductory reminder of properties of the current-current (CC)
correlation function in vacuum (for a review see e.g. ref. \cite{11}).
The vector mesons are seen as resonances in the current-current (CC)
correlation function,
\begin{equation}
  \Pi_{\mu\nu}(q)=i\int d^4x \: e^{iq\cdot x}\langle
  0|{\cal T}j_{\mu}(x) j_{\nu}(0)|0\rangle  ,
 \label{2.1}
\end{equation} 
where ${\cal T}$ denotes the time-ordered product and $j_{\mu}$ is the
electromagnetic current which can be decomposed as
\begin{equation}
  \label{2.1b}
  j_{\mu}= j_\mu^{(\rho)}+j_\mu^{(\omega)}+j_\mu^{(\phi)}
\end{equation}
into vector currents specified by their quark
content:
\begin{eqnarray}
\label{2.2a}
  j^{(\rho)}_\mu&=&\frac{1}{2}(\bar{u}\gamma_{\mu}u-\bar{d}\gamma_{\mu}d),\\
\label{2.2b}
  j^{(\omega)}_\mu&=&\frac{1}{6}(\bar{u}\gamma_{\mu}u+\bar{d}\gamma_{\mu}d),\\
  j^{(\phi)}_\mu&=&-\frac{1}{3}(\bar{s}\gamma_{\mu}s).
\label{2.2c}
\end{eqnarray}
Current conservation implies the following transverse tensor structure:
\begin{equation}
  \Pi_{\mu\nu}(q)=\left(g_{\mu\nu}-\frac{q_{\mu}q_{\nu}}{q^2}\right)\Pi (q^2),
\label{2.3}
\end{equation}
with the scalar CC correlation function
\begin{equation}
  \Pi (q^2)=\frac{1}{3} g^{\mu\nu}\Pi_{\mu\nu}(q).
\label{2.4}
\end{equation}
Its imaginary part is proportional to the cross
section for $e^+ e^- \to$ hadrons: 
\begin{equation}
 R(s)=\frac{\sigma(e^+ e^-\to \rm hadrons )}{\sigma (e^+
  e^-\to \mu^+ \mu^-)}=-\frac{12 \pi}{s} \Im \Pi(s),
\label{2.6}
\end{equation}
where $\sqrt{s}$ is the total c.m. energy of the lepton pair and $\sigma (e^+
e^-\to \mu^+ \mu^-)= 4 \pi \alpha^2/3 s$ with $\alpha=e^2/4 \pi=1/137$. The
vector mesons can be distinguished by looking at different hadronic channels
carrying the respective flavour (isospin) quantum numbers (see
eqs.(\ref{2.2a}-\ref{2.2c})). The corresponding correlation functions are
denoted by $\Pi^{(I=0)}$, $\Pi^{(I=1)}$ and $\Pi^{(\phi)}$. G-parity demands
that the isovector current involving the $\rho$ meson can only couple to even
numbers of pions (see the data in fig. 1a). Similarly, the annihilation of
$e^+e^-$ into odd numbers of pions determines the isoscalar current involving
the $\omega$ meson, as can be seen in Figure 1b. Isospin violating processes
are small but visible, such as the $\rho \omega$ mixing seen in the pion
formfactor. The situation for the $\phi$ meson is more involved.  The OZI rule
would prohibit decays into pions. However, a violation of this rule by about
five percent is observed in the three-pion channel.  (Some experimental data
near the $\phi$ resonance have been left out in Fig. 1b to emphasize the
$\omega$-contribution given by the solid line). The $\phi$ meson decays mainly
into OZI allowed channels such as $K^+K^-$ or $K_LK_S$ (see Fig. 1c), but these
channels also couple to the $\rho$ and $\omega$ mesons.  The measured cross
section of the annihilation into kaons includes an interference between all
three vector mesons. (For more details see ref.\cite{12}).  Nevertheless the
$\phi$ meson still dominates the data in the $e^+e^- \to K \bar{K}$ channels.

\subsection{Vector Meson Dominance update}
In the region $\sqrt{s} \;\lsim \; 1 \; \GeV$ vector meson dominance (VMD)
describes the experimental data of $e^+e^-$ annihilation into hadronic
channels very well. We briefly summarize here the results of ref. \cite{12}
based on an effective Lagrangian which combines chiral dynamics and vector
meson dominance. This approach starts out from a "bare", stable vector meson
with a mass $\stackrel{ \rm o }{m}_V$ slightly larger than the physical mass.
Its propagator is $\stackrel{\rm o}{D}_V\!(q^2)=(q^2-\stackrel{ \rm o
  }{m}_V^2)^{-1}$. The coupling to the pseudoscalar mesons turns $\stackrel{\rm
  o}{D}_V$ into the physical vector meson propagator in vacuum,
\begin{eqnarray}
  \label{2.7}
  D_V(q^2) &=& \nonumber  \stackrel{\rm o}{D}_V\!(q^2)+\stackrel{\rm o}{D}_V\!(q^2)
  \Pi_V(q^2) D_V(q^2) \\
&=& \frac{1}{q^2-\stackrel{ \rm o }{m}_V^2-\Pi_V(q^2)},
\end{eqnarray}
with the complex vector meson self energy $\Pi_V$. The decay width of a given
vector meson is related to the imaginary part of $\Pi_V$ by
\begin{equation}
  \label{2.10}
  m_V \Gamma_V = - \Im \Pi_V (q^2=m_V^2),
\end{equation} 
where  $m_V^2= \stackrel{ \rm o }{m}_V^2+ \Re \;
\Pi_V(q^2=m_V^2)$ represents the squared physical vector meson mass.
\subsubsection{The $\rho$ meson}
For example, the coupling of
the $\rho$ meson to the two-pion continuum leads to
\begin{equation}
  \label{2.8}
  \Im \, \Pi_\rho (q^2) = -\frac{g^2}{48 \pi} q^2 \left(1-\frac{4
  m_\pi^2}{q^2}\right)^{\frac{3}{2}} \Theta ( q^2 - 4 m_\pi^2 ) \; ,
\end{equation}
where $g=6.05$ is the $\rho \pi \pi$ coupling constant and $m_\pi=139.6 \, \MeV$
the charged pion mass. The real part of $\Pi_\rho$ can be derived from a
dispersion relation. It has the form
\begin{equation}
  \label{2.9}
  \Re \: \Pi_\rho (q^2)  =  c \;q^2-\frac{g^2}{24\pi^2} \left[ q^2 \;
  {\cal G} (q^2,m_\pi^2) -4 m_\pi^2 \right]\, ,
\end{equation}
with the function ${\cal G}(q^2,m^2)$ given explicitly in ref.\cite{12}. The
subtraction constant $c=0.16$ is determined by a fit to the isovector p-wave
$\pi \pi$ scattering phase shift, which also fixes the bare rho meson mass
$\stackrel{ \rm o }{m}_\rho=0.81 \, \GeV$. Note that in this approach the
physical rho mass in vacuum, $m_\rho=0.77 \, \MeV$, is already shifted downward
from the "bare" mass $\stackrel{ \rm o }{m}_V$. One may think of this "bare"
rho meson as being a quark-antiquark bound state before turning on its strong
coupling to the $\pi \pi$ continuum. ( In the $1/N_c$ counting scheme, the bare
rho meson mass is leading order in $N_c$, whereas the coupling to the $\pi \pi$
continuum via the complex self-energy $\Pi_V$ introduces $1/N_c$ corrections.)

\subsubsection{The $\omega$ meson}
The relation (\ref{2.10}) can be used to determine the imaginary part of the omega meson
self energy by the decay into three pions ($\pi^+,\:\pi^-,\:\pi^0$), as follows:
\begin{equation}
  \label{2.10a}
  \Im \Pi_\omega(q^2)=-\frac{q^2}{192 \pi^3} \int \int dE_+\; dE_- \big[
\vec{p}_-^{\,2} \vec{p}_+^{\,2}-(\vec{p}_- \cdot \vec{p}_+)^2\big] \; |F_{3\pi}|^2.
\end{equation}
Here $E_{\pm}$ and $\vec{p}_\pm$ are the energies and momenta of the two
charged pions in the final state, and the integration extends over the
kinematically allowed region in the ($E_+,\;E_-$) plane. The amplitude
$F_{3\pi}$ represents the sum of the direct $\omega \to 3 \pi$ decay and the
Gell-Mann, Sharp, Wagner (GSW) process in which the omega meson converts first
into $\pi \rho$, followed by the decay of the $\rho$ meson into two pions:
\begin{equation}
  \label{2.10b}
  F_{3\pi}=  \frac{0.18}{f_\pi^3}+ 2.4 \frac{g}{f_\pi} \sum_i D_\rho (q-p_i).
\end{equation}
Here we have introduced the pion decay constant $f_\pi=92.4 \, \MeV$, and
$D_\rho$ is the rho meson propagator. The four-momentum $p_i$ refers to that of
the first pion produced in the primary $\omega \to \pi \rho$ transition of the
GSW process, and the index $i$ denotes the three possible charges of that
primary pion. The determination of the coupling constants in eq.(\ref{2.10b})
is described in detail in ref.\cite{12}. The real part of $\Pi_\omega$ can in
principle be obtained by a dispersion relation. However, we choose to absorb
this real part into the physical $\omega$ meson mass. This is justified since
the momentum dependence of $\Re \; \Pi_\omega$ does not influence the results
in any significant way \cite{12}, given the small omega meson width.

\subsubsection{The $\phi$ meson}
The self energy of the $\phi$ meson consists of three parts:
\begin{equation}
  \label{2.10c}
  \Pi_\phi = \Pi_{\phi \to K^+ K^-} + \Pi_{\phi \to K^0_L K^0_S}+ \Pi_{\phi \to
  3 \pi}\; .
\end{equation}
The last term violates the OZI rule, but despite the small $\phi \omega $ mixing
angle $\theta \simeq 0.05$ it contributes about fifteen percent of the total $\phi$ decay
width. It is given by eq.(\ref{2.10a}) multiplied by the squared $ \phi \omega$ mixing angle. The
self energy parts coming from the decays into $K\bar{K}$ channels are of the
same form as the $\pi \pi$ self energy of the
$\rho$-meson and connected with the latter through SU(3) symmetry:
\begin{eqnarray}
  \label{2.10d}
  \Re \Pi_{\phi \to K^+K^-} (q^2)&  = &c_K \, q^2 -\frac{g^2}{48 \pi^2} \left[q^2 {\cal G}
 (q^2,m_{K^+}^2)- 4 m_{K^+}^2\right]\; , \\
\Im \;\Pi_{\phi \to K^+K^-} (q^2) &  = &- \frac{g^2}{96\pi} q^2 \left( 1 -
\frac{4m_{K^+}^2}{q^2} \right) ^\frac{3}{2} \; \Theta (q^2-4m_{K^+}^2)\; , 
\end{eqnarray}
For $\Pi_{\phi \to K^0_L K^0_S}$ the
same expressions hold with the charged kaon mass $m_{K^+}$ replaced by
$m_{K^0}$. The subtraction constant $c_K=0.11$ has been determined in Ref.\cite{12}. Comparing the theoretical widths with empirical ones we find that
SU(3) symmetry works well and gives the same universal coupling constant $g$
within ten percent \cite{12} for all couplings of the vector and pseudoscalar
meson octets.
 
\subsubsection{Current Correlation functions and VMD}

We can now establish the relation between the reduced electromagnetic CC
correlation function $\Pi$ and the vector meson self-energies $\Pi_V$. Our
starting point is again the effective Lagrangian of the coupled system of
pseudoscalar mesons, vector mesons and photons as described in ref. \cite{12}.
This approach includes not only a direct conversion of a photon into a vector
meson proportional to $e q^2/\stackrel{ \rm o}{g}_V$, but also a term which
involves the photon interaction with pseudoscalar meson currents. The couplings
$\stackrel{ \rm o}{g}_V$ are related through SU(3) symmetry as $ \stackrel{ \rm
  o}{g}_\rho \,:\, \stackrel{ \rm o}{g}_\omega \,:\, \stackrel{ \rm o}{g}_\phi
= 1:3:\frac{-3}{\sqrt{2}}$. The same relation holds for $g_V$, the strong couplings,
where we identify $g_\rho \equiv g$. The result for $\Pi(q^2)$, expressed in
terms of $\Pi_V(q^2)$, when iterating the vector meson self-energies to all
orders is as follows:
\begin{equation}
  \label{2.12a}
 \Pi(q^2)= \sum_V \frac{1}{g_V^2} \, \left(\Pi_V(q^2)+\frac{\gamma^2_V(q^2)
 q^4}{q^2-\stackrel{ \rm o }{m}_V^2-\Pi_V(q^2)} \right).
\end{equation}
This can easily be verified by direct diagrammatic analysis (see Fig. 2) using
the framework of ref. \cite{12}. Here $\gamma_V(q^2)$ represents the photon -
vector meson vertex with inclusion of vertex corrections due to pion or kaon
loops:
\begin{equation}
  \label{2.12b}
  \gamma_V(q^2)=a_V-\frac{\Pi_V(q^2)}{q^2}.
\end{equation}
In case of the $\rho$ meson we have $a_\rho=g/\stackrel{ \rm o}{g}_\rho \simeq 1.1$, the
ratio of the $\rho\pi\pi$ and $\rho\gamma$ couplings. In the limiting case of
``complete'' VMD where all of the photon-hadron interaction is transmitted via
vector mesons, we have $a_V \equiv 1$. In the $\omega$ and $\phi$ channels the
information about the direct $\gamma \leftrightarrow 3\pi$ vertex is very
limited and we therefore assume $a_\omega=a_\phi=1$. This gives an excellent
description of the $e^+e^- \to \pi^+\pi^0\pi^-$ cross section \cite{12}.

The imaginary part of $\Pi$ can now be easily evaluated:
\begin{equation}
  \label{2.12c}
  \Im \Pi (q^2)= \sum_V \frac{\Im \Pi_V(q^2)}{g_V^2} \left| F_V(q^2) \right|^2,
\end{equation}
where 
\begin{equation}
  \label{2.12d}
   F_V(q^2) = \frac{(1-a_V)\, q^2-\stackrel{ \rm o }{m}_V^2}{q^2-\stackrel{ \rm o
   }{m}_V^2-\Pi_V(q^2)}\, .
\end{equation}
For example, in case of the rho meson, $F_\rho(q^2)$ is simply identified with the pion form factor:
\begin{equation}
  \label{2.12e}
    F_\pi(q^2)=\frac{\left(1- \frac{g}{\stackrel{ \rm o
   }{g}_\rho}\right)q^2-\stackrel{ \rm o }{m}_\rho^2}{q^2-\stackrel{ \rm o
   }{m}_\rho^2-\Pi_\rho(q^2)}\, ,
\end{equation}
where again $g$ ($\stackrel{ \rm o }{g}_\rho$) is the coupling of the $\rho$
meson to the pion (photon) and $\Pi_\rho$ is the $\rho$ meson self energy
specified in eqs.(\ref{2.8},\ref{2.9}).

\subsubsection{High energy limit}

In the high energy region the measured correlation function approaches the
asymptotic plateau predicted by perturbative QCD:
\begin{equation}
  R(s)= -\frac{12 \pi}{s} \Im \Pi(s) =\sum_V d_V \left( 1+\frac{\alpha_S}{\pi} \right) \Theta (s- s_V)
\label{2.13}
\end{equation}
for $s$ larger than a characteristic scale $s_V$, where $d_\rho=3/2$,
$d_\omega=1/6$ and $d_\phi=1/3$. This behaviour is clearly seen in the
$\rho$-meson channel where the low mass (two-pion) region of the spectral
function is evidently given by $\rho$ meson dominance in the form of
eqs.(\ref{2.12c}) and (\ref{2.12e}). For $s > s_\rho \simeq 1.5 \, \GeV^2$
contributions from 4 and 6 pions take over and reach the asymptotic plateau
above $s \sim 2 \, \GeV^2$.

\subsection{QCD sum rule analysis}
The basic idea of QCD sum rules is to match two representations of $\Pi(q^2)$
in the region of large spacelike $Q^2=-q^2 \gg 1 \; \GeV$ for each one of the
channels $V=\rho$, $\omega$, $\phi$ (we drop the channel index and always refer
to a specific channel in this section). One representation is in terms of a
twice subtracted dispersion relation:
\begin{equation}
  \Pi(q^2)=\Pi (0)+c \, q^2+\frac{q^4}{\pi}\int ds\frac{\Im\Pi (s)}{s^2(s-q^2
  -i\epsilon)}, 
\label{2.14}
\end{equation}
where the vanishing photon mass in the vacuum requires $\Pi(0)=0$. The second
representation is in terms of the operator product
expansion (OPE):
\begin{equation}
  \label{2.14a}
  \frac{12 \pi}{Q^2} \Pi(q^2=-Q^2) =
  \frac{d}{ \pi} \left[-c_0 \ln{\left(\frac{Q^2}{\mu^2}\right)}+ \frac{c_1}{Q^2}+\frac{c_2}{Q^4}+\frac{c_3}{Q^6}+... \right]
\end{equation}
(at a scale $\mu$ which is commonly chosen around 1 GeV). Here the coefficients $c_{1,2,3}$ incorporate the
non-perturbative parts coming from the QCD condensates. In the isovector channel one
finds
\begin{eqnarray}
  \label{2.15}
 c^\rho_0&=& 1+\frac{\alpha_S(Q^2)}{\pi} ,\hspace*{1cm} c^\rho_1=-3(m_u^2+m_d^2), \\
c^\rho_2&=& \frac{\pi^2}{3} \langle 0|\frac{\alpha_S}{\pi} {\cal G}^{\mu \nu}{\cal
G}_{\mu \nu}|0\rangle + 4 \pi^2 \langle0| m_u u\bar{u}+ m_d d\bar{d}
|0 \rangle,\\ \label{2.15d}
c^\rho_3&=&-4 \pi^3 \left[\langle 0|\alpha_S(\bar{u}\gamma_\mu \gamma_5 \lambda^a
u-\bar{d}\gamma_\mu \gamma_5 \lambda^a d)^2 |0\rangle \right.\nonumber \\&& \left.+\frac{2}{9} \langle0|
\alpha_S(\bar{u}\gamma_\mu \lambda^a u+\bar{d}\gamma_\mu  \lambda^a d) \sum_{q=u,d,s}
\bar{q} \gamma^\mu \lambda^a q |0\rangle \right].  
\end{eqnarray}
Here $\alpha_S=4 \pi / (\, b \ln{(Q^2/\Lambda_{QCD}^2)})$ is the running
coupling constant (with $\Lambda_{QCD}=140\, \MeV$ and $b=11-(2/3)N_f=9$). We
can set $c_1$ equal to zero because its contribution is proportional to the
squared mass of the light quarks and therefore negligible. For the isoscalar
channel there is only a sign change between the up and down quark condensates
of the first term in coefficient $c_3$.

The situation is different in the $\phi$ meson channel because of the large
strange quark mass $m_s$.  The coefficient $c_1^\phi$ is not negligible any longer
and $c_2^\phi$ is larger than $c_2^{\rho,\omega}$. One finds \cite{15}
\begin{eqnarray}
  \label{2.15a}
 c^\phi_0&=& 1+\frac{\alpha_S(Q^2)}{\pi} ,\hspace*{1cm} c^\phi_1=-6m_s^2, \\
\label{2.15b}
c^\phi_2&=& \frac{\pi^2}{3} \langle 0|\frac{\alpha_S}{\pi} {\cal G}^{\mu \nu}{\cal
G}_{\mu \nu}|0\rangle + 8 \pi^2 \langle0| m_s s\bar{s}
|0 \rangle,\\ \label{2.15c}
c^\phi_3&=&-8 \pi^3 \left[\langle 0|\alpha_S(\bar{s}\gamma_\mu \gamma_5 \lambda^a
s)^2 |0\rangle \right.\nonumber \\&& \left.+\frac{4}{9} \langle0|
\alpha_S(\bar{s}\gamma_\mu \lambda^a s) \sum_{q=u,d,s}
\bar{q} \gamma^\mu \lambda^a q |0\rangle \right].  
\end{eqnarray}
For the gluon condensate we use 
\begin{equation}
  \label{2.16}
  \langle 0| \frac{\alpha_S}{\pi} {\cal G}^2|0\rangle=1.2 \cdot 10^{-2} \, \GeV^4,
\end{equation}
the value determined from charmonium sum rules \cite{15}.  The four-quark
condensates which contribute to $c_3$ are usually factorized by assuming that
vacuum saturation holds, i.e. that they are proportional to
$\langle0|\bar{q}q|0 \rangle^2$. Then the coefficient $c_3$ is the same for all
three channels. While this is a valid approximation in the large $N_c$ limit
\cite{16}, possible deviations from vacuum saturation must be taken into
account. We introduce a parameter $\kappa>1$ for this purpose and write:
\begin{equation}
  \label{2.17}
  c_3=-\frac{896}{81}\kappa \; \pi^3 \; \alpha_S \langle0| \bar{q}q|0 \rangle^2 .
\end{equation} 
Here it is still assumed that the four-quark condensates with any two different
flavours vanish. While this is not rigorously correct, the error can be
estimated from $\rho \omega$ mixing \cite{15}. Its smallness suggests that this
assumption may be appropriate.  With the commonly accepted values
$\langle0|\bar{u}u|0 \rangle \simeq \langle0|\bar{d}d|0 \rangle \simeq
\langle0|\bar{s}s|0 \rangle\equiv \langle0|\bar{q}q|0 \rangle \simeq(-250\,
\MeV)^3$ one needs $\kappa=2.36$ in order to obtain the same value of $c_3$ as
originally proposed by Shifman et al. \cite{15}. Note that Hatsuda et al.
\cite{17} use a larger value $ \langle0|\bar{q}q|0 \rangle=(-280 \, \MeV)^3$ in
their approach. We also mention here that strongly increased values of $c_3$ as
suggested in refs.  \cite{18,19} would not be compatible with our analysis.

The QCD sum rule approach compares the dispersion relation of eq.(\ref{2.14})
on the left side with its OPE representation (eq.(\ref{2.14a}), right side). We
use our calculated
\begin{equation}
  \label{2.18}
  R^{(V)}(s) =-\frac{12 \pi}{s\, g_V^2} \Im \Pi_V(s) |F_V(s)|^2 \Theta(s_V-s)+d_V \left(1+\frac{\alpha_S}{\pi}\right) \Theta(s-s_V)
\end{equation}
to determine the left side. Since the OPE on the right side of eq.(\ref{2.14a}) is truncated at
order $Q^{-6}$ we apply the usual Borel transformation \cite{14,15} in
order to improve convergence:
\begin{equation}
  \label{2.19}
  \hat{f}(\M^2)= \lim_{Q^2 \to \infty,\; n \to
  \infty, \atop (Q^2/n \equiv \M^2 ={\rm const.})} \frac{1}{(n-1)!} (Q^2)^n \left(
  - \frac{d}{d  Q^2} \right)^n f(Q^2).
\end{equation}
After Borelization the QCD sum rule for each $R^{(V)}$ becomes
\begin{equation}
  \label{2.20} 
   \frac{1}{\M^2\, d} \left[12 \pi^2 \Pi(0) +\int_0^{\infty}ds\, R (s)
  e^{-s/ \M^2} \right] = c_0+\frac{c_1}{\M^2}+\frac{c_2}{\M^4}+\frac{c_3}{2 \M^6}.
\end{equation}
Given $R^{(V)}(s)$ of eq.(\ref{2.18}) this represents a consistency condition
which should hold independently of the Borel mass parameter $\M$ within the
range  of applicability of the QCD sum rules, i.e. for $\M$ larger than a
certain minimal value. Note again in the vacuum $\Pi(0)$ vanishes, but we have
kept it here since it will not vanish in matter. 

In Fig. 3 we compare the ``left side'' of eq.(\ref{2.20}) calculated from
eq.(\ref{2.18}) with the ``right side'' as a function of the Borel mass $\cal
M$. There is excellent agreement in all three channels between both sides of
the sum rule, down to $\M \simeq 0.8 \, \GeV$ where the convergence of the OPE
series (\ref{2.14a}) starts to break down and higher order terms become
important. We have checked that our condensates agree with the "optimal"
values, using the fit program MINUIT. It determines the condensate parameters
to be
\begin{eqnarray}
c^{\rho,\omega}_2&=&0.04\, \GeV^4,\\
c^{\rho,\omega}_3&=&-0.071 \, \GeV^6
\end{eqnarray}
 together with $s_{\rho,\omega}=1.56 \, \GeV^2$ and
\begin{eqnarray}
c^\phi_1&=&0.07\, \GeV^2,\\
c^\phi_2&=&-0.10\, \GeV^4,\\
c^\phi_3&=&-0.07 \, \GeV^6,
\end{eqnarray}
taking $s_\phi=2.2 \, \GeV^2$.

We also confirm previous results \cite{15} which use the simplified ansatz for the spectra,
\begin{equation}
  \label{2.21}
  R^{(V)} ={\cal F}_V \; \delta (s-m_V^2)+d_V \left(1+\frac{\alpha_S}{\pi}\right) \Theta(s-s_V)
\end{equation}
with ${\cal F}_\rho=9\,{\cal F}_\omega=2.4$, $m_{\rho,\omega}=0.77\, \GeV$,
$s_{\rho,\omega}=1.5\, \GeV^2$ and ${\cal F}_\phi=0.79$, $m_\phi=1.02\, \GeV$,
$s_\phi=2.2\, \GeV^2$. Including the vector meson decay widths does not make a
substantial difference, at least not in vacuum: after taking the Borel
transform their effects are suppressed as $(m_V \Gamma_V/\M^2)^2$. This
explains why the large difference of the $\rho$ and $\omega$ meson decay widths
does not affect the OPE. However, for fine tuning purposes it is essential to
use the complete spectrum (\ref{2.18}). For the in-medium sum rule the use of a
realistic spectrum is mandatory as discussed in section \ref{5}. We also note
that the approximate degeneracy of the $\rho$ and $\omega$ meson masses emerges
naturally in the OPE. This is one of the traditional successes of QCD sum
rules.

\section{\label{3} Correlation functions in baryonic matter}

We now replace the vacuum $|0\rangle$ by a state $|\Omega(\rho) \rangle$ of
matter with given baryon density $\rho$. The CC correlation function to leading
order in $\rho$ has been written down before \cite{17,20}. In this chapter we
briefly summarize the steps which lead to this first order term, and we also
describe its iteration to all orders in density. The basic input
into this scheme is the forward Compton amplitude for scattering of a timelike,
virtual photon on the nucleon. Our task is to calculate this amplitude using an
effective Lagrangian approach. The resulting spectral function in matter will
then be further analyzed using a QCD sum rule with in-medium condensates.

\subsection{Basics}

In the present paper we restrict ourselves to the case of matter at temperature
$T=0$. The matter may be moving with a constant four-velocity $v^{\mu}$. We
choose a frame of reference in which matter as a whole is at rest by setting
$\vec{v}=0$. The four-momentum $q^{\mu}=(\omega,\vec{q}\, )$ that enters in the
CC correlation function is from now on understood to be taken in this
particular frame. The generalization of the CC correlation function to the case
of baryonic matter is
\begin{equation}
\label{3.1}
  \Pi_{\mu\nu}(\omega ,\vec{q};\rho)=i\int\limits^{+\infty}_{-\infty}dt\int
  d^3 x\: e^{i\omega t-i\vec{q}\cdot\vec{x}}
  \left<\Omega(\rho)\left|{\cal T}j_{\mu}(t,\vec{x})j_{\nu}(0)
  \right|\Omega (\rho)\right>.
\end{equation}
For the vacuum case ($\rho=0$) we have seen that at squared virtual photon
masses $q^2=\omega^2-\vec{q}^{\; 2} \, \lsim \, 1 \, \GeV^2$, the spectrum of
$\Pi$ starts at $q^2=4 m_\pi^2$ in the isovector channel, at $q^2=9 m_\pi^2$ in
the isoscalar channel and at $q^2=4 m_K^2$ in the $\phi$ channel (apart from
small $\phi \to 3\pi$ contribution which also starts at $q^2=9 m_\pi^2$).
This spectrum displays the vector mesons as the dominant doorway states which
decay into pseudoscalar mesons. In matter we have new processes involving
nucleons, such as
\begin{equation}
  \label{3.2}
  \gamma^*N \to \pi N,\, \pi \pi N,\,K\Lambda,\,K\Sigma,\hspace{1cm} {\rm etc.}
\end{equation}
which start at $q^2 \geq m_\pi^2$ or at the threshold for kaon production. In
addition there are nuclear particle-hole excitations at very low energies which
need not be considered here.

For a nuclear Fermi gas with equal number of protons and neutrons, Fermi
momentum $p_F$ and density $\rho=2p_F^3/(3\pi^2)$, the density dependent matrix
element $\langle\Omega(\rho)|\hat{ \cal O }| \Omega(\rho) \rangle$ of any
operator $\hat{\cal O}$ is given by \cite{21}:
\begin{equation}
  \label{3.3}
  \langle\Omega(\rho)|\hat{ \cal O }| \Omega(\rho) \rangle=\langle 0 |\hat{
  \cal O }| 0 \rangle+4 \int_{|\vec{p}\,| \leq p_F} \frac{d^3p}{(2\pi)^3}
  \frac{M}{E(\vec{p}\,)}   \langle N(\vec{p}\,) |\hat{ \cal O }| N(\vec{p}\,)
  \rangle+... \, ,
\end{equation}
where $|N(\vec{p}\,) \rangle$ is a proton or neutron state (including
spin) with momentum $\vec{p}$ and the normalization
\begin{equation}
 \label{3.4}
\langle N(\vec{p}\,)|N(\vec{p}\,')\rangle=(2\pi)^3 \frac{E(\vec{p}\,)}{M} \delta^3 (\vec{p}-\vec{p}\,'),  
\end{equation}
with $E(\vec{p}\,)=\sqrt{\vec{p}^{\; 2}+M^2}$. The dots in eq.(\ref{3.3})
indicate higher orders in density.

When applied to the CC correlation function in matter, eq.(\ref{3.3}) becomes
\begin{equation}
  \label{3.5}
  \Pi_{\mu\nu}(\omega ,\vec{q};\rho)=\Pi^{\rm vac}_{\mu\nu}(\omega ,\vec{q}\, )+4\int_{|\vec{p}\,|\le p_F}\frac{d^3 p}{(2\pi)^3
  }\frac{M}{E(\vec{p}\,)}T_{\mu\nu}(\omega ,\vec{q};\vec{p}\,)+... \, ,
\end{equation}
where the first term $\Pi_{\mu \nu}^{\rm vac}= \Pi_{\mu \nu}(\rho=0)$ is just the vacuum CC correlation
tensor (\ref{2.1}) discussed previously, and
\begin{equation}
  \label{3.6}
  T_{\mu\nu}(\omega ,\vec{q};\vec{p}\,)=i\int^{+\infty}_{-\infty}dt\int d^3 x
  \: e^{i\omega t-i\vec{q}\cdot\vec{x}}\langle N(\vec{p}\,)\left|{\cal T}j_{\mu}(t,\vec{x}
  )j_{\nu}(0)\right|N(\vec{p}\,)\rangle
\end{equation}
is the forward Compton scattering tensor for a virtual photon of energy
$\omega$ and momentum $\vec{q}$ on a free nucleon. The tensor $T_{\mu \nu}$ has
the standard decomposition
\begin{equation}
  \label{3.7}
  T_{\mu \nu}= \left(\frac{q_{\mu}q_{\nu}}{q^2}-g_{\mu\nu}\right)T_1
  +\left(p_{\mu}-\frac{p\cdot q}{q^2}q_{\mu}\right) \left(p_{\nu}-\frac{p\cdot
  q}{q^2}q_{\nu}\right) \frac{T_2}{M^2},
\end{equation}
which defines the amplitudes $T_1$ and $T_2$. It is often more useful to
introduce the amplitudes

\begin{eqnarray}
  \label{3.8}
  T^{(T)}&=&T_1, \\
 \label{3.8b}
  T^{(L)}&=& T_1-\left[ 1-\frac{(p\cdot q)^2}{p^2 \, q^2} \right] T_2,
\end{eqnarray}
and reexpress $T_{\mu \nu}$ as
\begin{equation}
  \label{3.9}
  T_{\mu \nu}= -(P_L)_{\mu \nu} \, T^{(L)} -(P_T)_{\mu \nu} T^{(T)} ,
\end{equation}
where the matrices $P_L$ and $P_T$ act as projectors on longitudinal and
transverse parts with the properties: $ P_L^2 = P_L $, $  P_T^2= P_T$
and $ P_L \, P_T = P_T \, P_L = 0$. For a nucleon at rest with
$p^{\mu}=(M,\vec{p}=0)$ the tensors $P_L$ and $P_T$ are given by
\begin{equation}
  \label{3.10}
   P_L= -\left[  \begin{array}{ccc} 
        \frac{\vec{q}^{\, 2}}{q^2} & \frac{\omega q^j}{q^2} \\
        \frac{\omega q^i}{q^2} & \frac{q^i q^j \omega^2}{q^2 \vec{q}^{\,
           2}} \end{array} \right]\;
 , \hspace{1cm} 
  P_T =  -\left[  \begin{array}{cc} 
           0 & 0 \\
           0 & \delta^{ij}-\frac{q^i q^j}{\vec{q}^{\; 2}} \end{array} \right].
\end{equation}

\subsection{Low density limit}

 At low baryon density $\rho$ the $\vec{p}$ dependence of the Compton tensor in
 eq.(\ref{3.5}) can be neglected and we have 
\begin{eqnarray}
\label{3.11}
  \Pi_{\mu \nu}(\omega ,\vec{q};\rho) &\simeq& \Pi^{\rm vac}_{\mu \nu}(q)+\rho
  \, T_{\mu
  \nu}(\omega ,\vec{q};\vec{p}=0)+...\,\\ \nonumber 
 & = & \Pi^{\rm vac}_{\mu \nu}(q)-\rho \left[ (P_L)_{\mu \nu}
 T^{(L)}(\omega,\vec{q},\vec{p}=0)+(P_T)_{\mu \nu}
 T^{(T)}(\omega,\vec{q},\vec{p}=0) \right]+...\, .
\end{eqnarray}
We introduce the notation $T_{\mu \nu}(\omega,\vec{q}\, ) \equiv T_{\mu
  \nu}(\omega,\vec{q};\vec{p}=0)$ etc.  from here on. The reduced CC
correlation function $\Pi (\omega ,\vec{q};\rho)=g_{\mu \nu} \Pi^{\mu \nu}/3$
analogous to eq.(\ref{2.4}) becomes
\begin{eqnarray}
\label{3.12}
  \Pi (\omega ,\vec{q};\rho) &= &\Pi^{\rm vac}(q)-\rho \left(\frac{1}{3} T^{(L)}(\omega,\vec{q}\, ) +
  \frac{2}{3} T^{(T)}(\omega,\vec{q}\, )\right) \nonumber \\
&=& \Pi^{\rm vac}(q)-\rho \left(T_1 ( \omega
  ,\vec{q}\, )+\frac{\vec{q}^{\,2}}{3q^2} T_2 (\omega,\vec{q}\, ) \right),
\end{eqnarray}
to leading order in $\rho$. These expressions have a direct analogue when
looking at the leading density dependence of the vector meson self-energy
tensor:
\begin{equation}
\label{3.13}
  \Pi_V^{\mu \nu} (\omega ,\vec{q};\rho)= \Pi_V^{{\rm vac} \; \mu \nu
  }(q^2)+\rho \, T_{V N}^{\mu \nu} (\omega, \vec{q}\, )+...\, ,
\end{equation}
where $\Pi_V^{{\rm vac} \, \mu \nu}=\Pi_V^{ \mu \nu}(\rho=0)$ is the vacuum
contribution and $T_{V N}^{\mu \nu} (\omega,\vec{q}\, )$ denotes the
(off-shell) vector meson-nucleon amplitude analogous to the Compton tensor $T^{\mu
  \nu}$.  With the help of the projectors $P_T$ and $P_L$ we can separate this
self energy into two parts:
\begin{equation}
\label{3.14}
  \Pi_V^{\mu \nu} (\omega ,\vec{q};\rho)= [\Pi_V^{{\rm vac} } (q^2)-\rho
  T_{V N}^{(L)}] (P_L)_{\mu \nu}+ [\Pi_V^{{\rm vac} } (q^2)-\rho T_{V N}^{(T)}] (P_T)_{\mu \nu} ,
\end{equation}
where $\Pi_V^{\rm vac}$ is the scalar self-energy function in the vacuum and $T_{V N}^{(L,T)}$ are the longitudinal and
transverse $V N$ scattering amplitudes analogous to the Compton amplitudes
$T^{(L,T)}$ of eqs.(\ref{3.8},\ref{3.8b}).

\subsection{Vector meson propagator and current correlation function in medium}
Using the projectors $P_L$ and $P_T$ it is now a straightforward exercise to
decompose the tensor structure of the in-medium vector meson propagator. We find 
\begin{equation}
  \label{3.15}
  D^{\mu \nu}_V(\omega,\vec{q};\rho) =
  \frac{(P_L)^{\mu \nu}}{q^2-\stackrel{ \rm o }{m}_V^2-\left(\Pi^{\rm
  vac}_V(q^2)-\rho \, T_{V N}^{(L)}\right)}+
  \frac{(P_T)^{\mu \nu}}{q^2-\stackrel{ \rm o }{m}_V^2-\left(\Pi^{\rm
  vac}_V(q^2)-\rho \, T_{V N}^{(T)}\right)}.
\end{equation}
The leading medium corrections, iterated to all orders, are therefore simply
obtained by the replacement $\Pi^{\rm vac}_V(q^2) \to \Pi^{\rm
vac}_V-\rho T_{V N}^{(L,T)}$ in the transverse and longitudinal parts,
respectively.

The structure seen in eq.(\ref{3.15}) translates directly into the in-medium CC
correlation function (\ref{3.1}). Using diagrammatic techniques analogous to
those in the vacuum one derives:
\begin{equation}
    \label{3.16}
    \Pi^{\mu \nu} (\omega,\vec{q};\rho)= \Pi^{(L)}(\omega,\vec{q};\rho)
    (P_L)^{\mu \nu} + \Pi^{(T)} (\omega,\vec{q};\rho) (P_T)^{\mu \nu},
\end{equation}
with the longitudinal and transverse parts
\begin{eqnarray}
  \label{3.17}
  &&\Pi^{(L,T)} (\omega,\vec{q};\rho) =\nonumber  \\ &&\sum_V \frac{1}{g_V^2} \left(\Pi_V^{\rm vac} (q^2)-\rho
  T_{V N}^{(L,T)}(\omega,\vec{q}\, ) +\frac{\left[\gamma_{V}^{(L,T)}(w,\vec{q};\rho)\right]^2 q^4}{q^2-\stackrel{ \rm o
  }{m}_V^2-\Pi^{\rm vac}_V(q^2)+\rho \, T_{V N}^{(L,T)}(\omega,\vec{q}\,
  )}\right) \quad
\end{eqnarray}
and
\begin{equation}
  \label{3.18}
  \gamma_{V }^{(L,T)}(\omega,\vec{q};\rho)= a_V-\frac{\Pi^{\rm vac}_V-\rho \, T_{V N}^{(L,T)}(\omega,\vec{q}\, )}{q^2}.
\end{equation}
This result generalizes eqs.(\ref{2.12a},\ref{2.12b}). The imaginary parts of
$\Pi^{(L)}$ and $\Pi^{(T)}$ become:
\begin{equation}
  \label{3.19}
   \Im \Pi^{(L,T)}=\sum_V \frac{1}{g_V^2} Im \left(\Pi^{\rm vac}_V-\rho T_{V
  N}^{(L,T)}\right) \left|\frac{(1-a_V) q^2-\stackrel{ \rm o
  }{m}_V^2}{q^2-\stackrel{ \rm o }{m}_V^2-\Pi^{\rm vac}_V+\rho
  \, T_{V N}^{(L,T)} }\right|^2. 
\end{equation}
This is the in-medium generalization of eq.(\ref{2.12c}), with $\Pi_V^{\rm
vac}$ replaced everywhere by $\Pi^{\rm vac}_V-\rho \,T_{V N}^{(L,T)}$.

It is instructive to recover the low-density limit for which one finds:
\begin{equation}
  \label{3.20}
   \Pi_{\mu \nu} (\omega,\vec{q};\rho) =  \Pi_{\mu \nu}^{\rm vac
  }(q^2)-\rho \sum_V \frac{1}{g_V^2} \, F_V(q^2) \left[T_{V N}^{(L)} (P_L)_{\mu \nu} + T_{V
  N}^{(T)} (P_T)_{\mu \nu} \right] F_V(q^2)+... 
\end{equation}
where $F_V(q^2)$ are the  form factors given in eq.(\ref{2.12d}). We can now identify the
relationship between the Compton scattering amplitude and the $V N$
amplitude by comparison with eq.(\ref{3.11}):
\begin{equation}
  \label{3.21}
  T_{\mu \nu}= \sum_V  \frac{1}{g_V^2} F_V \left[T_{V N}^{(L)} (P_L)_{\mu \nu} + T_{V
  N}^{(T)} (P_T)_{\mu \nu} \right] F_V
\end{equation}
or
\begin{equation}
  \label{3.22}
  T^{(L,T)}= \sum_V \frac{F_V^2}{g_V^2}  T_{V N}^{(L,T)} .
\end{equation}
In the limiting case of a real photon with $q^2 = \omega^2-\vec{q}^{\; 2} =0$ we
have $F_V(q^2=0) = 1$ and find the simpler relation
\begin{equation}
  \label{3.23}
   T^{(L,T)}(\omega=|\vec{q}| )= \sum_V \frac{T_{V N}^{(L,T)}(\omega=|\vec{q}| )}{g_V^2}.
\end{equation}
Furthermore, $\Pi^{\rm vac}(q^2=0)=0$ and $\Im T^{(L)}=0$ for real photons so
that we obtain, to leading order in the density $\rho$:
\begin{equation}
  \label{3.24}
   \Im \Pi_{\mu \nu} (\omega=|\vec{q}\,|;\rho) = -\rho \, \Im
   T^{(T)}(\omega=|\vec{q}\,|) \, (P_T)_{\mu \nu},
\end{equation}
and therefore
\begin{equation}
  \label{3.25}
  \sum_V  \frac{\Im T_{V N}^{(T)}(\omega=|\vec{q}\, |)}{g_V^2}=\frac{\omega}{4 \pi \alpha} \,  \sigma_{\gamma N}(\omega),
\end{equation}
where $\sigma_{\gamma N}$ is the total photon-nucleon cross section averaged
over proton and neutron and we have used the
optical theorem
\begin{equation}
  \label{3.26}
  \sigma_{\gamma N}(\omega)=e^2 \Im T^{(T)}(\omega=|\vec{q}\, |)/\omega.
\end{equation}

\subsection{Vector mesons at rest}
From now on we focus on the simpler case $\vec{q}=0$ in which the vector
mesons are at rest. Note that this is also the situation corresponding to
back-to-back emission of an electron-positron pair. In
this case we have 
\begin{equation}
  \label{3.27}
  T(\omega)\equiv -\frac{1}{3} g_{\mu \nu} T^{\mu \nu
  }(\omega,\vec{q}=0)=T_1(\omega,\vec{q}=0)=T^{(T)}(\omega,\vec{q}=0)=T^{(L)}(\omega,\vec{q}=0).
\end{equation}
The reduced CC correlation function in the low density limit (eq.(\ref{3.11}))
now becomes
\begin{equation}
\label{3.28}
  \Pi (\omega ,\vec{q}=0;\rho) = \Pi^{\rm vac}(\omega^2)-\rho \, T (
  \omega)+... \, ,
\end{equation}
and for the vector meson self energy we find
\begin{equation}
\label{3.29}
  \Pi_V (\omega ,\vec{q}=0;\rho)= \Pi_V^{\rm vac}(\omega^2)-\rho \, T_{V N} (
  \omega)+... \, ,
\end{equation}
where $T_{V N}=T_{V N}^{(L)}= T_{V N}^{(T)}$ denotes the (off-shell) vector meson-nucleon
amplitude for $\vec{q}=0$, analogous to the Compton amplitude $T(\omega^2)$. 
The full in-medium propagator (\ref{3.15}), for a vector meson at rest, simplifies
to $D_V^{00}=D_V^{0j}=D_V^{i0}=0$ and
\begin{equation}
  \label{3.30}
  D^{ij}_V (\omega,\vec{q}=0;\rho) =
  \frac{-\delta^{ij}}{\omega^2-\stackrel{ \rm o }{m}_V^2-\left(\Pi^{\rm
  vac}_V(\omega^2)-\rho \,T_{V N}\right)}.
\end{equation}
The reduced CC correlation function in matter which follows from eq.(\ref{3.16})
is now:
\begin{equation}
  \label{3.31}
   \Pi (\omega,\vec{q}=0,\rho) = \sum_V \frac{1}{g_V^2} \left(\Pi_V^{\rm vac} (\omega^2)-\rho
  T_{V N} +\frac{\gamma_V^2(\omega;\rho) \, \omega^4}{\omega^2-\stackrel{
  \rm o}{m}_V^2-\Pi^{\rm vac}_V(\omega^2)+\rho \, T_{V N}} \right),
\end{equation}
with
\begin{equation}
  \label{3.32}
    \gamma_{V}(\omega;\rho)= a_V-\frac{\Pi^{\rm vac}_V(\omega^2)-\rho \,T_{V N}(\omega)}{\omega^2}. 
\end{equation}
Its imaginary part enters directly in the back-to-back dilepton
production rate and it is also the key input for the ``left side'' of the QCD
sum rule: 
\begin{eqnarray}
  \label{3.33}
  && \Im \Pi(\omega,\vec{q}=0;\rho)=\nonumber \\ && \sum_V \frac{1}{g_V^2} \Im
  \left(\Pi^{\rm vac}_V(\omega^2)-\rho T_{V N}(\omega)\right)
  \left|\frac{\left(1- a_V\right)\omega^2-\stackrel{ \rm o
  }{m}_V^2}{\omega^2-\stackrel{ \rm o }{m}_V^2-\Pi^{\rm vac}_V(\omega^2)
  +\rho \, T_{V N}(\omega)}\right|^2. \quad  
\end{eqnarray}
We have thus established the relationship between the isovector CC correlation
function in matter and the (complex) off-shell vector meson-nucleon amplitudes. 

\section{\label{4} Vector meson-nucleon scattering amplitudes}
\subsection{Effective Lagrangian}
In order to determine the low energy $VN$ scattering amplitudes, we use
the hadronic effective Lagrangian of ref.\cite{12} which combines chiral SU(3)
dynamics with VMD. This approach has been successfully applied for vacuum
processes. The extended Lagrangian
incorporates nucleons, hyperons (and also $\Delta$'s in intermediate states
) with pseudovector meson-baryon interactions: 
\begin{equation}
  \label{4.1}
   {\cal L}_{\Phi B} = F \,{\rm tr}\left( \bar{B} \gamma_\mu
   \gamma_5 [u^\mu,B]\right)+D\, {\rm tr}\left( \bar{B} \gamma_\mu
   \gamma_5 \{u^\mu,B\}_+\right),  
\end{equation}
where $B$ ($\Phi$) represent the SU(3) matrix fields of the baryon (pseudoscalar meson) octet and 
\begin{equation}
  \label{4.2}
  u^\mu= -\frac{1}{2f_\pi} \left(\partial^\mu \Phi -ie [Q,\Phi] A^\mu \right).
\end{equation}
This effective Lagrangian has been tested successfully in the SU(3) coupled
channels approach of refs.\cite{22}. Here $Q$ denotes the quark charge matrix,
$A^\mu$ is the photon field and $f_\pi=92.4 \, \MeV$ is the pion decay
constant. The second part of $u^\mu$ gives rise to the Kroll-Ruderman couplings
of the photon with the interacting system of baryons and pseudoscalar mesons. The
parameters $F\simeq0.51$ and $D\simeq0.75$ satisfy the constraint $g_A=F+D=1.26$.
The vector meson-baryon interactions are then introduced by the minimal
coupling scheme which substitutes $ eQ A^\mu$ by the vector field $g V^\mu/2$,
where
\begin{equation}
  \label{4.2b}
    V^\mu  \equiv  \left(  \begin{array}{ccc} 
         \rho^\mu+\omega^\mu & 0 & 0 {\rule[-3mm]{0mm}{8mm}} \\
             0 & -\rho^\mu+\omega^\mu & 0  {\rule[-3mm]{0mm}{8mm}}\\
             0   & 0 & \sqrt{2} \phi^\mu \end{array} \right) \; .
\end{equation}
In the Kroll-Ruderman part this leads to
\begin{equation}
  \label{4.3}
   {\cal L}_{V \Phi B} = \frac{ig}{4 f_\pi} \left( F \, {\rm tr}( \bar{B} \gamma_\mu
   \gamma_5 [[V^\mu,\Phi],B])+D\,{\rm tr}( \bar{B} \gamma_\mu
   \gamma_5 \{[V^\mu,\Phi],B\}_+ ) \right).  
\end{equation}
The direct coupling of the photon to the baryons becomes:
\begin{equation}
  \label{4.4}
   {\cal L}_{\gamma B} =  e\left({\rm tr}( \bar{B} \gamma_\mu [Q,B]
   )-{\rm tr}( \bar{B} \gamma_\mu B ){\rm tr}(Q) \right)
   A^\mu .
\end{equation}
The second term vanishes, because the trace of the $u-,\,d-$ and $s-$quark
charge matrix is zero. When determining the direct vector meson-baryon coupling
by the replacement $ eQ A^\mu \to g V^\mu/2$ this term survives and we find:
\begin{equation}
  \label{4.5}
   {\cal L}_{V B}^{(1)} = \frac{g}{2} \left({\rm tr}( \bar{B} \gamma_\mu [V^\mu,B] )-{\rm tr}( \bar{B} \gamma_\mu B ){\rm tr}(V^\mu) \right) .
\end{equation}
This form implies the SU(3) symmetry relation $g \equiv g_{\rho N} =
\frac{1}{3} g_{\omega N}=-\frac{\sqrt{2}}{3} g_{\phi N}$.  We also include
corrections due to anomalous $VN$ tensor couplings with $\kappa_\rho=6.0$ and
$\kappa_\omega=0.1$:
\begin{equation}
  \label{4.5b}
  {\cal L}_{V N}^{(2)} = \frac{g\kappa_\rho}{4 M_N } \bar{N} \vec{\tau}
  \sigma_{\mu \nu}N \partial^\mu \vec{\rho}^{\,\nu}+ \frac{g\kappa_\omega}{4 M_N } \bar{N}  \sigma_{\mu
  \nu}N \partial^\mu  \omega^\nu.
\end{equation}
In the strict SU(3) limit we set $\kappa_\phi=0$ and recognize that there
is no direct $\phi$ meson-nucleon coupling. The $\phi N$ interaction in this
channel arises entirely from kaon loops.

The interaction vertices obtained from the effective Lagrangian are summarized
in Fig. 4. The minimal coupling scheme for the vector mesons makes sure that
we correctly approach the limit $\omega \to 0$ keeping $\vec{q}=0$.

Using this effective Lagrangian we now calculate the $\rho$, $\omega$ and
$\phi$ meson-nucleon scattering amplitudes $T_{VN}(\omega)$, with the vector
mesons at rest ($\vec{q}=0$).  The imaginary parts of these scattering
amplitudes are then evaluated using standard Cutkosky rules. The real parts are
determined by a subtracted dispersion relation at $\vec{q}=0$:
\begin{equation}
  \label{4.7}
  \Re T_{V N} (\omega) = l_V+ \frac{\omega^2}{\pi}
  {\cal P} \int_{0}^\infty du^2 \frac{\Im T_{V N} (u)}{u^2
  (u^2-\omega^2)}.
\end{equation}
The subtraction constants $l_V$ are fixed by the Thomson limit at $\omega=0$.
Using eq.(\ref{3.23}) we have $l_\rho=l_\omega/9=-g^2/(4M_N)$ and $l_\phi=0$.
Evidently, processes with low thresholds dominantly influence the low energy
region of the real and imaginary parts of $T_{VN}$. The convergence of the
dispersion integral is slow, however, because $\Im \, T_{VN}$ increases
linearly with energy. Therefore its high energy part also influences $\Re \,
T_{VN}$.

Our hadronic effective Lagrangian is expected to be valid up to a
momentum and energy scale of order 1 GeV and we expect formfactor suppressions
at higher energies, in particular when the mesons are off-shell. Indeed, in the
vacuum the VMD approach seizes to be valid at a scale $s=s_V$, and perturbative
QCD takes over for $s > s_V$. We introduce formfactors for vector meson-baryon
couplings:
\begin{equation}
  \label{4.7b}
  F_{VB}(k^2)=\frac{\Lambda_V^2-m_V^2}{\Lambda_V^2-k^2},
\end{equation}
where $k$ is the four momentum, $m_V$ the mass of the vector meson, and
$\Lambda_V$ is the
cutoff parameter. We choose $\Lambda_V=1.6 \, \GeV$, values close to those
commonly used in Boson exchange $NN$ potentials. At the pseudovector pion-nucleon
vertices we employ the empirical axial form factor
 \begin{equation}
  \label{4.7c}
  G_A(k^2)=\frac{g_A}{(1-k^2/\Lambda^2_A)^2},
\end{equation}
with $\Lambda_A \simeq 1 \, \GeV$ \cite{36} and assume that a slightly larger
$\Lambda_A^{\Lambda,\Sigma}=1.2 \, \GeV$ applies for kaon couplings to baryons.
We will discuss the sensitivity of the results to variations of the momentum
scales $\Lambda$.

\subsection{ The $\rho N$ amplitude}
For the $\rho$ meson the dominant contributions to the imaginary part of the
scattering amplitude come from the inelastic processes $\rho N \to \pi N$,
$\rho N \to \pi \Delta \to \pi \pi N$ and $\rho N \to \omega N \to \pi \pi \pi
N$. The corresponding one-loop diagrams are illustrated in Fig. 5. The
propagating baryon in the loops of this figure can either be a nucleon or a
$\Delta$ isobar.  In the limit of large baryon mass only the diagrams, Figs.
5(a,c,e,f,m), survive. First we omit the form factor $G_A(k^2)$ and calculate
the imaginary parts of the loops using the bare coupling $g_A$. For the $\pi N$
loops we get
\begin{equation}
  \label{4.8}
   \Im T_{\rho N}^{(\pi N)}(\omega) = g_A^2 \, {\cal H}(\omega,0,m_\pi)
\end{equation}
where the function $\cal H$ is given (for $\omega>0$) by:
\begin{eqnarray}
  \label{4.8a}
  {\cal H}(\omega,\Delta,m)& = & \nonumber \frac{g^2}{24 \pi
  f_\pi^2} \left[\frac{\sqrt{(\omega-\Delta)^2-m^2}}
  {(\omega^2-2\omega \Delta)^2} \left(3 \omega^4-4 \omega^2 m^2+4 m^4
  +\Delta (8 \omega m^2-12 \omega^3)\right. \right. \\
  \nonumber  && \hspace*{1.2cm}\left. \left.+\Delta^2 (16 \omega^2-8m^2 )-8 \Delta^3 \omega +4 \Delta^4
  \right)\right.\\  && \hspace*{1.2cm} \left.  -\frac{ \Delta (\omega^2-4
m^2)^\frac{3}{2}}{2 \omega \, (\omega^2-4 \Delta^2)^2} \left(3
\omega^2-8m^2-4 \Delta^2 \right) \right] .
\end{eqnarray}
If we include diagram 5 k, which does not contribute to the imaginary part of
$T_{\rho N}$, we can also evaluate the real part in the heavy baryon limit
since all divergences cancel due to current conservation. We then find for the
$\pi N$ loop contribution to $\Re\,T_{\rho N}$ in this limit :
\begin{equation}
  \label{4.9}
  \Re \, T_{\rho N}^{(\pi N)}(\omega) = \frac{g^2 g_A^2 }{24 \pi
  f_\pi^2}\left[\frac{2m_\pi^3}{\omega^2}-\frac{3 m_\pi}{2}+ \left(3 -\frac{4
  m_\pi^2}{\omega^2}+\frac{4 m_\pi^4}{ \omega^4} \right) \left(\sqrt{m_\pi^2-\omega^2} \, \Theta(m_\pi-\omega)-m_\pi
  \right) \right] .
\end{equation}
This corresponds to the result obtained by using eq.(\ref{4.8}) as input in
the dispersion relation (\ref{4.7}). Of course we have to take relativistic
corrections and the axial form factor into account. Calculating the $\pi N$
loops fully relativistically and using $G_A(k^2)$ given in eq.({\ref{4.7c})
instead of $g_A$ we end up with $\Im T_{\rho N}^{(\pi N)}(\omega)$ as shown
by the short-dashed line of Fig. 7a.

Note that for the $\pi N$ loops the function ${\cal H}(\omega,0,m_\pi)$ greatly
simplifies. The situation is more involved when we consider the $\pi \Delta$
loops. We find in the heavy baryon limit
 \begin{equation}
\label{4.8b}
  \Im T_{\rho N}^{(\pi \Delta)}(\omega) = 2 \,g_A^2 \, {\cal H}(\omega,M_\Delta-M_N,m_\pi).
\end{equation}
Here $M_\Delta=1.232 \, \GeV$ is the $\Delta$-isobar mass. A fully relativistic
calculation would require using the Rarita-Schwinger formalism which introduces
ambiguities. Here we only take the axial form factor into account by replacing
$g_A$ by a dipole $\pi N \Delta$ transition formfactor $G_A^{(\Delta)}(k^2)$
with $\Lambda_A^{(\Delta)}= \Lambda_A$. Adding this result to the $\pi N$ loops
contribution gives rise to the long-dashed curve in Fig. 7a.

For the high energy part of the spectrum we have found the box diagram 5(m) to be very
important. Its fully relativistic evaluation gives 
\begin{equation}
  \label{4.8c}
  \Im T_{\rho N}^{(\pi \omega)}(\omega)= g_A^2\, {\cal I}(\omega,m_\omega),
\end{equation}
 where
\begin{equation}
  \label{4.8d}
  {\cal I}(\omega,m_V)=  \frac{g_{\omega \rho \pi}^2 }{48
  \pi f_\pi^4} \frac{\omega^2\, \vec{k}^{\, 3} \left[2\vec{k}^{\,2} M_N^2-M_N
  (k_0-\omega) \left[(k_0-\omega)^2-\vec{k}^{\,2} \right] \right]} 
{ M_N (M_N+\omega)(m_\pi^2-m_V^2+2  k_0 \omega-\omega^2)^2} 
\end{equation}
Here $g_{\omega \rho \pi}=1.2$ and we use $|\vec{k}|=\sqrt{\lambda (M_N+\omega,m_V,M_N)} /(2(M_N+\omega
))$ and $k_0=\sqrt{m_V^2+\vec{k}^{\,2}}$. The K{\"a}llen function $\lambda$ is
defined by
\begin{equation}
  \label{4.8e}   
  \lambda(x,y,z)=(x^2-(y+z)^2)\, (x^2-(y-z)^2).
\end{equation}
Including the axial formfactor and adding this to the contributions of the $\pi
N$ and $\pi \Delta$ loops gives rise to the solid curve in Fig. 7a.

One might expect that processes $\rho N \to K \Sigma, \, K \Lambda$ also
contribute to $T_{\rho N}$. We give the result from $K \Sigma$ and $K \Lambda$
loops in the heavy baryon limit, setting $m_K=m_{K^+}=m_{K^0}$:
\begin{equation}
  \label{4.8f}
  \Im T_{\rho N}^{(K \Lambda,\,  K \Sigma)}(\omega) = \frac{3}{8} (D-F)^2 \,{\cal
  H}(\omega,M_\Sigma-M_N,m_K)+ \frac{1}{24} (D+3F)^2\, {\cal
  H}(\omega,M_\Lambda-M_N,m_K)
\end{equation}
Here $M_\Lambda=1.116 \, \GeV$ and $M_\Sigma=1.191 \, \GeV$ are the masses of
$\Lambda$ and $\Sigma$ hyperons. Note that SU(3) factors reduce this amplitude
already by about one order of magnitude with respect to the $\pi N$ loops, and a further reduction comes from the much smaller phase space and
relativistic corrections. We can therefore safely neglect those contributions. 

The real part of the $\rho N$ scattering amplitude is then determined  by using
$\Im T_{\rho N}$ (solid line of Fig. 7a) as input for the dispersion
relation (\ref{4.7}). The result is shown if Fig. 8a. The decomposition of $\Re
T_{\rho N}$ into subprocesses is also given in this figure.

\subsection{The $\phi N $ amplitude}
For the $\phi$-meson the same contributions from the $K \Lambda$ and $K \Sigma$
loops as for the $\rho$-meson appear, but now larger by a factor of two:
\begin{equation}
  \label{4.10}
  \Im T_{\phi N}(\omega) = \frac{3}{4} (D-F)^2\, {\cal
  H}(\omega,M_\Sigma-M_N,m_K)+ \frac{1}{12} (D+3F)^2 \,{\cal
  H}(\omega,M_\Lambda-M_N,m_K)
\end{equation}
Note that from SU(3) symmetry the $\phi N$ amplitude is already four times
smaller than the $\rho N$ amplitude with inclusion of $\pi N$ loops only. A
further reduction comes from the smaller phase space because of the higher $KN$
threshold. Therefore the imaginary part and consequently also the real part
(shown in Fig. 8c) of the $\phi N$ amplitude is much reduced as compared to the
$\rho N$ case. This is evident by comparing Figs. 7a and c. The smallness of
$\Im T_{\phi N}$ implies that effects of regularization by meson-baryon
formfactors have little impact on the real part, at least in the region around
the physical $\phi$ mass. We show results in Fig. 8c with inclusion of a
kaon-baryon dipole formfactor using a cutoff
$\Lambda^{(\Lambda,\,\Sigma)}_A=1.2 \, \GeV$. We mention however that diagrams
such as Fig 5(l,m), with $K^*$ and $K$ mesons exchanged in the box are not taken
into account in the present calculation. Very little is known about such
processes, but they may not be negligible.

\subsection{The $\omega N$ amplitude}
The Kroll-Ruderman term and the coupling to two pseudoscalar mesons do not
exist for the $\omega$ meson. Consequently no such diagrams as in Figs.
5(a,b,c,e,f,h,i) contribute to the imaginary part of the $\omega N$ amplitude.
The diagrams, Figs. 5(d,g,j) vanish in the heavy baryon limit. They contribute
very little to the scattering amplitude even with the large coupling
$g_\omega=17$. The dominant contributions come from processes where the
$\omega$ turns into a $\pi \rho$ system which then interacts with the nucleon.
The corresponding box diagrams, Fig. 5(m) (with the role of $\rho$ and $\omega$
interchanged) give:
\begin{equation}
  \label{4.11}
  \Im T_{\omega N}^{(1)}(\omega) = 3 \, g_A^2 \, {\cal I}(\omega,m_\rho),
\end{equation}
where ${\cal I}(\omega,m_V)$ is defined in eq. (\ref{4.8d}).  At low energy the
box diagram Fig. 5(l) becomes important since the rho mesons in the box
interact strongly with the nucleons due to the large anomalous tensor coupling
with $\kappa_\rho=6$. We find for this term:
\begin{equation}
  \label{4.11b}
  \Im T_{\omega N}^{(2)}(\omega) = \frac{g_{\omega \rho \pi}^2 g^2
  (1+\kappa_\rho)^2}{128\, \pi f_\pi^2 }\frac{|\vec{l}|\, 
  (\omega^2-m_\pi^2)^2 \,\omega^2 \,(4 M_N^2-m_\pi^2+4 M_N \omega
  +\omega^2)}{(M_N+\omega)^4\,(m_\rho^2-m_\pi^2-\omega^2+2 \omega l_0)^2}. 
\end{equation}
using $|\vec{l}|=\sqrt{\lambda (M_N+\omega,m_\pi,M_N)}/(2(M_N+\omega))$ and
$l_0=\sqrt{m_\pi^2+\vec{l}^{\, 2}}$ where $\lambda$ is the K{\"a}llen function
given in eq.(\ref{4.8e}).

Including also the contributions from vertex corrections in which the $\omega$
couples on one side directly to the nucleon and on the other side to the $\pi
\rho$ system, we show the resulting imaginary part of the $\omega N$ amplitude
in Fig. 7b replacing $g_A$ by the axial form factor and multiplying
$\kappa_\rho$ by the monopole form factor (\ref{4.7b}). At low energies $\Im
T_{\omega N}$ is small but rises strongly at higher energies. The high energy
behaviour is sensitive to the meson-baryon form factors. This unavoidable model
dependence also translates into $\Re \, T_{\omega N}$ (see Fig. 8b) when using
the calculated $\Im T_{\omega N}$ in the dispersion relation (\ref{4.7}). We
should also point out, however, that the behaviour of $\Im \, T_{\omega N}$ can
at least partly be checked against data. For example, the mechanisms that
determine the long-dashed curve in Fig. 7b also enter in the s-wave part of the
$\pi N \to \omega N$ cross section. We find indeed satisfactory results for
this cross section close to threshold. 

\subsection{Effective vector meson - nucleon scattering length}

In order to gain further insights into the basic features of the $VN$
amplitudes taken at $\vec{q}=0$, it is instructive to introduce complex
effective scattering lengths,
 \begin{equation}
   \label{4.12}
    a_{VN}=\frac{M_N}{4 \pi (M_N+m_V)}T_{V N}(\omega=m_V),
 \end{equation}
defined for ``on-shell'' vector mesons at rest. These scattering length
parameters should of course be interpreted with care since the vector mesons
themselves are unstable. Nevertheless the real and imaginary parts of $a_{VN}$
give a direct impression of the strong interaction shifts and widths that the
vector mesons would experience in nuclear matter at low density.

The following effective scattering lengths are obtained for our standard
scenario, with input specified in section 4:
\begin{eqnarray}
\label{4.13}
a_{\rho N} &= &(0.04  +i\, 1.62 )\,\fm ,\\
a_{\omega N} &=& (3.34 +i \,2.1 )\, \fm, \\ 
\label{4.14}
a_{\phi N} &=& (-0.01  +i \, 0.08 )\, \fm.
\end{eqnarray}
We observe that the $\rho N$ scattering length is strongly dominated by its
imaginary part which describes the inelastic $\rho N \to \pi N,\, \pi \pi N$
etc. channels, whereas $\Re\, a_{\rho N}$ is persistently small, more than an
order of magnitude smaller than $\Im \, a_{\rho N}$. In essence, this $a_{\rho
  N}$ causes a strong inelastic broadening, but almost no mass shift in matter.
This is at striking variance with almost all previous analyses which have
simply ignored the large imaginary part of $a_{\rho N}$.

The situation is quite different for the $\omega N$ scattering length. There
one finds a large (attractive) real part of $a_{\omega N}$ which is mainly driven
by the processes, Fig. 5(l,m). Thus the $\omega$ meson mass is indeed expected to
be shifted downward in matter. At the same time one observes a substantial
inelastic broadening.

Finally, the comparatively weak $\phi N$ interaction is reflected in the
small $a_{\phi N}$, with almost vanishing real part and a moderate $\Im \,
a_{\phi N}$ which comes primarily from the inelastic channels $\phi N \to K
\Lambda,\, K \Sigma$. 

The results, eqs.(\ref{4.13}-\ref{4.14}), are of course subject to some model
dependence related to the form factors at the meson-baryon vertices that enter
into loop diagrams, Fig. 5. Changes by 20\% of the cutoffs $\Lambda$ in these
form factors induce variations at the 30\% level in the real and imaginary
parts of $a_{VN}$. For example, increasing the cutoff $\Lambda_A^{(\Delta)}$ of
the dipole $\pi N \Delta$ transition form factor from 1 GeV to 1.3 GeV will
change the $\rho N$ scattering length to $a_{\rho N}=(-0.02+i1.4)$fm. It is
thus uncertain whether the very small $\rho$ meson mass shift in matter is
upward or downward, but the primary feature of a large $\Im \, a_{\rho N}$
remains in all cases. We have also examined corrections due to $\pi \pi$
rescattering in the basic $\rho N \to \pi\pi N$ box diagram (see Fig. 6) and
found them to be small. Such mechanisms produce attraction, but their effects
on the $\rho$ mass shift in matter are still weak (less than 5\% at $\rho=\rho_0$)

\section{\label{5} Spectral densities and QCD sum rules in medium}
\subsection{ In-medium spectral functions}

Combining our results for the $V N$ scattering amplitudes with the relations
developed in chapter 3 we are now able to compute the in-medium correlation
function (\ref{3.29},\ref{3.31}). We insert $T_{VN}$ into eq.(\ref{3.31}) and
display the spectral function in the form $-(12 \pi /\omega^2)\, \Im\, 
\Pi(\omega,\rho)$ for convenience, the one corresponding to
eq.(\ref{2.6}). This is done separately in each of the vector meson channels for
different baryon densities $\rho$.

\subsubsection{Isovector spectral function}
In the isovector channel (see Fig. 9a) we observe the following characteristic
features \cite{23} as compared to the vacuum:

- The imaginary part of the in-medium rho meson self energy is strongly
  increased by the large $\Im \, T_{\rho N}$. This leads to a huge width in the CC
  correlation function, with the consequence that the rho meson resonance
  broadens and flattens.

- The spectrums starts now already at $\omega=m_\pi$, the threshold energy at
  which single pion production becomes possible. This generates a strong
  continuum in the correlation function below the $\rho$ resonance. We recall
  that the vacuum spectrum starts at a threshold $\omega=2 m_\pi$ coming
  from  $\rho \to \pi^+ \pi^-$ decay.

- The shift of the $\rho$ meson pole position in matter is only marginal, given
  the small real part of the $\rho N$ scattering length. The in-medium rho
  meson mass is determined by
  \begin{equation}
    \label{5.1}
    m^2_\rho(\rho)=\stackrel{ \rm o }{m}_\rho^2+\Re \left[\Pi^{\rm
    vac}_\rho(m^2_\rho)-\rho T_{\rho N}(m_\rho^2)\right] .
  \end{equation}
  With our standard scenario  we find
  \begin{equation}
    \label{5.2}
    m^2_\rho(\rho)=m_\rho^2(0)\left (1-0.005\frac{\rho}{\rho_0}\right), 
  \end{equation}
a very small shift at nuclear matter density, $\rho_0=0.17
\fm^{-3}$. 

It is instructive to study the real part of the in-medium meson propagator,
\begin{equation}
\label{5.3}
 D_\rho (\omega,\vec{q}=0;\rho) =
  \frac{1}{\omega^2-\stackrel{ \rm o }{m}_\rho^2-\left(\Pi^{\rm
  vac}_\rho(\omega^2)-\rho \,T_{\rho N}(\omega)\right)}.
\end{equation}
The zero of $\Re D_\rho $ determines the in-medium mass. The result is shown in
Fig. 10a. One clearly observes that, with rising density, the rho meson has
little chance to survive as a quasi-particle. The large inelastic width in
matter diffuses the well developed quasi-particle pole structure seen in the
vacuum. The actual downward shift of strength in the spectrum in matter is
caused entirely by the strong increase of the imaginary part of the in-medium
self energy.

\subsubsection{Isoscalar spectral function}

For the $\omega$ meson the situation is different (see Fig. 9b). It starts out
as a narrow resonance in vacuum. Both real and imaginary parts of $T_{\omega
  N}$ are large. The $\omega$ meson mass shifts downward and its width
increases by an order of magnitude at $\rho=\rho_0$ as compared to the free
width, but it survives as a quasi-particle in matter, at least up to densities
$\rho \simeq \rho_0$. This is clearly seen in the real part of the in-medium
$\omega$ meson propagator (see Fig. 10b), the zero of which moves downward with
increasing density while the pronounced resonance pattern still persists.

\subsubsection{The $\phi$ spectral function}

Because of the relatively weak $\phi N$ interaction, the $\phi$ meson peak
moves very little (see Fig. 9c). At $\rho=\rho_0$ its width increases by about
one order of magnitude over its very small decay width in free space due to the
opening of the inelastic channels $\phi N \to K \Lambda, K \Sigma$. The
spectrum now start at the threshold for those processes, but the well developed
$\phi$ resonance remains clearly visible at $\rho=\rho_0$.

\subsection{QCD sum rules in baryonic matter}

Our calculated in-medium CC correlation function at $\vec{q}=0$ is now used as
input for the dispersion relation which represents the left side of the QCD sum
rule at finite baryon density. We then examine the consistency with the
operator product expansion, the right hand side of this sum rule.

As in the vacuum case described in section \ref{2}, it is useful to work with
the quantities
\begin{equation}
  \label{5.4}
   R^{(V)}(\omega;\rho) =-\frac{12 \pi}{\omega^2\, g_V^2} \Im \, \Pi_V
   (\omega,\vec{q}=0,\rho) \, \Theta(s_V-\omega^2)+d_V \left(1+\frac{\alpha_S}{\pi}\right) \Theta(\omega^2-s_V)
\end{equation}
in each of the three vector meson channels ($V=\rho,\,\omega,\, \phi$). The
calculated spectrum $\Im \, \Pi_V$ (see eq.(\ref{3.33})) is valid up to some
(density dependent) scale $s_V(\rho)$ which still needs to be fixed. At higher
energies beyond $s_V$ the perturbative QCD result takes over; here we assume
that the coefficients $d_V$ do not change with density. After Borel
transformation one obtains the in-medium QCD sum rules in a form analogous to
eq.(\ref{2.20}): 
\begin{equation}
  \label{5.5} 
  \frac{1}{d_V \M^2} \left[12 \pi^2 \frac{T_{VN}(0) \rho}{g_V^2}
  +\int_0^{\infty}d\omega^2 R^{(V)}(\omega,\rho)\, e^{-\omega^2/ \M^2} \right]=
  \tilde{c}_0+\frac{\tilde{c}_1}{\M^2}+\frac{\tilde{c}_2(\rho)}{\M^4}+\frac{\tilde{c}_3(\rho)}{2 \M^6}.
\end{equation}
The first term on the left hand side corresponds to the Landau damping, the
vector meson analogue of the Thomson scattering limit, with $T_{VN}(0)=
g_V^2/4M_N$ for $V=\rho,\, \omega$. For the $\phi$ meson this term vanishes.

The coefficients $\tilde{c}_1$, $\tilde{c}_2$ and $\tilde{c}_3$ include the
density dependence of condensates which have already appeared in the vacuum sum
rule, as well as new condensates which only exist at finite baryon density. For
the isovector and isoscalar channels we have \cite{2}:
\begin{eqnarray}
  \label{5.7a}
\tilde{c}_0^{\rho,\omega}&=&c_0^{\rho,\omega}, \hspace*{1cm}\tilde{c}_1^{\rho,\omega}=0, \\
\label{5.7b}
\tilde{c}_2^{\rho,\omega}(\rho)&=& c_2^{\rho,\omega}(\rho)+2 \pi^2 A_1^{u+d}M_N\rho,\\
\label{5.7c}
\tilde{c}_3^{\rho,\omega}(\rho)&=& c_3^{\rho,\omega}(\rho)-\frac{10}{3} \pi^2 A_3^{u+d} M_N^3 \rho,
\end{eqnarray}
where $c_0^{\rho,\omega}$ are the same as in the vacuum. For
$c_2^{\rho,\omega}$ we use \cite{2}:
\begin{eqnarray}
  \label{5.8}
  c_2^{\rho,\omega}(\rho) &=& \frac{\pi^2}{3} \langle \Omega|\frac{\alpha_S}{\pi} {\cal G}^{\mu \nu}{\cal
G}_{\mu \nu}|\Omega\rangle \nonumber \\ &\simeq & \frac{\pi^2}{3} \langle 0|
\frac{\alpha_S}{\pi} {\cal G}^{\mu \nu}{\cal
G}_{\mu \nu}|0\rangle-\frac{8 \pi^2}{27} M_N^{(0)} \rho ,
\end{eqnarray}
where $M_N^{(0)}=750\,\MeV$ is chosen for the nucleon mass in the chiral limit \cite{37}. The
contribution from the chiral condensate, $\langle \Omega | m_u \bar{u}u+ m_d
\bar{d}d | \Omega \rangle=\langle0| m_u \bar{u}u+ m_d \bar{d}d|0\rangle+\rho \sigma_N$ with the nucleon sigma term $\sigma_N \simeq 45 \, \MeV$, is small and can be neglected.

The in-medium four-quark condensate determines $c_3^{\rho,\omega}(\rho)$. Its
density dependence is not very well under control and requires some
discussion. The basic question is to what extent ground state saturation, $\langle
\Omega|(\bar{q}q)^2|\Omega \rangle \simeq \langle \Omega| \bar{q}q | \Omega
\rangle^2$, is approximately valid in matter. This factorization of
$\langle(\bar{q}q)^2 \rangle$ holds in the mean field approximation where one
neglects any excitations that can be reached by acting with $\bar{q} q$ on the
ground state. In a nuclear medium as well as in the vacuum, scalar multi-pion
excitations of the ground state are expected to be relevant, however. By
analogy with eqs.(\ref{2.15d}) and (\ref{2.17}) we parameterize:
\begin{eqnarray}
  \label{5.9}
   \langle \Omega(\rho)|(\bar{q} \gamma_\mu \gamma_5\lambda^a
   q)^2|\Omega(\rho)\rangle &=& -\langle \Omega(\rho)|(\bar{q} \gamma_\mu
   \lambda^a q)^2|\Omega(\rho)\rangle \nonumber =\frac{16}{9}\kappa(\rho)
   \langle  \Omega(\rho)|\bar{q}q|\Omega(\rho)\rangle^2,
\end{eqnarray}
where the density dependent factor $\kappa(\rho)$ now represents the in-medium
deviation from simple factorization. Ground state saturation means $\kappa
\equiv 1$. We now assume that the leading density dependence comes from the
in-medium condensates $\langle  \Omega (\rho)|\bar{q}q|
\Omega(\rho)\rangle=\langle 0| \bar{q}q|0 \rangle + \sigma_N \,\rho /(m_u+m_d)$,
while the low-energy part of the spectrum of scalar excitations that enters in
$\kappa (\rho) >1$ is approximately independent of $\rho$ (see also
refs.\cite{3,24}). We therefore choose $\kappa(\rho) \equiv \kappa \simeq 2.36 $
as in the vacuum and find
\begin{equation}
\label{5.10}
   c^{\rho,\omega}_3=-\frac{896}{81}\kappa \; \pi^3 \; \alpha_S \left(\langle0|
   \bar{q}q|0 \rangle^2-\frac{2\sigma_N\, \rho }{m_u+m_d} \langle 0|\bar{q}q|0
   \rangle \right) 
\end{equation}
to leading order in density $\rho$. This may be a crude approximation which
overestimates the density dependence. At least the tendency for $\langle
(\bar{q}q )^2\rangle$ to decrease with increasing density is without question
\cite{25}, since the difference between vector and axial vector current
correlation functions, $|\langle| j_V j_V |\rangle - \langle| j_A j_A |\rangle 
|$, is proportional to $\langle \bar {q} q \rangle ^2$ and tends to zero at the
point of chiral restoration.

As pointed out in ref.\cite{2}, further in-medium contributions to
eqs.(\ref{5.7a}-\ref{5.7c}) arise from matrix elements of the form $\langle N|
\bar{q} \gamma_{\mu_1} D_{\mu_2}... D_{\mu_{ n+1}} q|N \rangle$ with gluon
fields entering in the gauge covariant derivatives $D_\mu$. These matrix
elements are proportional to the moments
\begin{equation}
  \label{5.11}
  A^q_{n}=2 \int_0^1 dx \, x^{n} [q(x)+\bar{q}(x)].
\end{equation}
of quark and antiquark distributions which can be determined, at some
renormalization scale of order 1 GeV, from deep-inelastic lepton scattering on
a nucleon. We use values as in ref.\cite{2} and set $A_1^{u+d}=0.9$,
$A_3^{u+d}=0.12$. Additional corrections of higher twist operators are smaller
than the twist 2 contributions \cite{26,27} and discussed in ref. \cite{17}.

For the $\phi$ meson the coefficient $c_2^\phi$ is dominated by the strange
quark condensate. Its in-medium change is given to leading order in density by
\begin{equation}
  \label{5.12}
  \langle \Omega(\rho)| \bar{s}s| \Omega(\rho) \rangle =  \langle 0| \bar{s}s|
  0 \rangle +y\, \frac{\rho \, \sigma_{N} }{m_u+m_d}
\end{equation}
with $y=\langle N| \bar{s}s |N\rangle /\langle N| \bar{u}u |N \rangle\simeq
0.2$. We then find
\begin{eqnarray}
  \label{5.13a}
\tilde{c}_0^{\phi}&=&c_0^{\phi}, \hspace*{1cm}\tilde{c}_1^{\phi}=c_1^\phi , \\
\label{5.13b}
\tilde{c}_2^{\phi}(\rho)&=& c_2^{\phi}(\rho)+4 \pi^2 A_1^{s}M_N\rho,\\
\label{5.13c}
\tilde{c}_3^{\phi}(\rho)&=& c_3^{\phi}(\rho)-\frac{20}{3} \pi^2 A_3^{s} m_N^3 \rho,
\end{eqnarray}
where $c_2^\phi (\rho)$ and $c_3^\phi(\rho)$ are obtained from  $c_{2,3}^\phi$
of eqs.(\ref{2.15b}-\ref{2.15c}) when replacing the vacuum $|0 \rangle$ by $|
\Omega (\rho)\rangle$. We use the values $A_1^{s}=0.05$ and
$A_3^{s}=0.002$ of the strange quark moments as in ref.\cite{2}. The resulting
density dependent changes of the right hand (OPE) side of the QCD sum rule for
the $\phi$ meson with respect to the vacuum are small.

In Figs. 11 we compare the ``left hand side'' with the ``right hand side'' of
eq.(\ref{5.5}) for the $\rho,\,\omega$ and $\phi$ channels at normal nuclear
matter density $\rho=\rho_0=0.17\, \fm^{-3}$, as a function of the Borel mass
$\M$. For the energy scales $s_V$ in eq.(\ref{5.4}) which separate the resonant
parts of the spectrum from the high-energy QCD continuum, we use
\begin{eqnarray}
\label{5.14a}
s_\rho&=&1.56\,  \GeV^2\,(1-0.1 \,\rho/\rho_0),\\
s_\omega&=&1.65 \, \GeV^2 (\,1-0.3\, \rho/\rho_0),\\ 
s_\phi&=&2.2 \, \GeV^2\,(1-0.01\, \rho/\rho_0),
\end{eqnarray}
in order to match the region of asymptotic Borel masses. 

The overall consistency between calculated in-medium spectral functions and the
density dependent operator product expansion is evidently quite remarkable. We
should compare Figs. 11a-c with the corresponding vacuum sum rule analysis,
Figs. 3a-c. Note that the difference between the $\rho=\rho_0$ and $\rho=0$
cases is large for the $\rho$ and $\omega$ meson channels, whereas it is
marginal for the $\phi$ channel.

\subsection{Discussion}
\subsubsection{Comparison with previous work}

Hatsuda and Lee \cite{2} have inserted the simplified parameterization
\begin{equation}
  \label{5.15}
  R^{(V)}(\omega,\rho)={\cal F}_V^* \delta(\omega^2-m_V^{*\,2}(\rho))+d_V
  \left( 1+\frac{\alpha_s}{\pi}\right) \Theta (\omega^2-s_V^*(\rho)) 
\end{equation}
in the ``left hand side'' of their in-medium QCD sum rule and fitted ${\cal F}_V^*,\,
m_V^*$ and $s_V^*$. Apart from their choice of $\kappa=1$, their ``right hand side''
is the same as eq.(\ref{5.5}). Their optimal fits gave
\begin{eqnarray}
   \label{5.16}
   \frac{m^*_{\rho,\, \omega}}{m_{\rho,\, \omega}}&=&  1-(0.16\pm0.6)
   \frac{\rho}{\rho_0} ,\\
   \sqrt{\frac{s^*_{\rho,\, \omega}}{s_{\rho,\,
   \omega}}}&=&1-(0.15\pm0.05)\frac{\rho}{\rho_0},\\ 
   \frac{{\cal F}^*_{\rho,\, \omega}}{{\cal F}_{\rho,\, \omega}}&=&1-(0.24\pm0.07)
   \frac{\rho}{\rho_0}. 
\end{eqnarray}
Here ${\cal F}_{\rho}=12 \pi^2 m_{\rho}^2/g_{\rho}^2$ is the vacuum pole strength for
the $\rho$ meson, and an analogous expression holds for the $\omega$ meson. For
the in-medium $\phi$ meson mass they find 
\begin{equation}
  \label{5.17}
  \frac{m^*_{\phi}}{m_{\phi}}=  1-(0.15\pm0.5) \,y\, \frac{\rho}{\rho_0},
\end{equation}
with $y=0.1-0.2$. Jin and Leinweber \cite{3} find relations consistent with
eqs.(\ref{5.16}-\ref{5.17}). If we use $\kappa=2$ the results
(\ref{5.16}-\ref{5.17}) change moderately within the errors indicated.

The fact that the in-medium masses $m^*_{\rho,\, \omega}$ both appear to
decrease by (10-20 \%) at $\rho=\rho_0$ in comparison with the vacuum masses
$m_{\rho,\, \omega}$ has been interpreted as an indication of BR scaling
\cite{1}. In contrast, we find in our explicit calculation that the low-energy
interaction of $\rho$ and $\omega$ mesons with surrounding matter differ
strongly. The naive ansatz (\ref{5.15}) remains reasonably justified in the
isoscalar channel. There we have shown that, while the effective width of the
$\omega$ meson at $\rho=\rho_0$ increases by about an order of magnitude over
its (small) free width, the in-medium $\omega$ meson can still be identified as
a quasi-particle with a well-defined effective mass that roughly resembles the
fit, eq.(\ref{5.16}). This is not the case, however, for the $\rho$ meson. The
inelastic collision broadening in the isovector channel is so strong that it
completely dominates the low-mass spectrum, whereas the pole mass of the rho
meson at $\rho=\rho_0$ changes very little. The average spectral mass $\bar{m}$
in this channel, defined by
\begin{equation}
  \label{5.18}
   \bar{m}^2(\rho)=\frac{\int_0^{1 \GeV^2} d\omega^2\, \omega^2 \,
   R^{I=1}(\omega,\rho) }{\int_0^{1 \GeV^2} d\omega^2 R^{I=1}(\omega,\rho)},
\end{equation}
drops roughly as $m^*_\rho$ in the parameterization (\ref{5.15}), but this can
obviously not be interpreted as a decreasing $\rho$ mass in matter, as we have
pointed out by analyzing the zero of the real part of the rho meson propagator
in Fig. 10a. 

Asakawa and Ko \cite{6} have used a more realistic $\rho$ meson spectral
function than the schematic zero width form (\ref{5.15}) and repeated the QCD
Sum rule analysis with inclusion of a density dependent rho meson width. They
have argued that a mass shift consistent with BR scaling still remains to be
added. Our
results for the $\rho$ meson do not support this conclusion.

The strong broadening of the $\rho$ meson spectral function in matter is also
found by Chanfray et al. \cite{35}. They include corrections of higher order in
density induced by
intermediate propagating pions. Our present calculations have focused on a
consistent calculation of vector
meson self-energies to leading order in density. We find that uncertainties in
the (small) $\rho$ meson mass shift related to vertex form factors and
high energy contributions to the spectral functions are in fact larger
than the effects of higher order corrections in pion propagators.

\subsubsection{Effective scattering amplitudes}

The main difference between the present approach and previous work is our
explicit calculation of the complex vector meson-nucleon amplitudes at
$\vec{q}=0$. In the low density limit the question about possible shifts of
vector meson masses reduces to a statement about the corresponding effective
$\rho N$, $\omega N$ and $\phi N$ scattering lengths. In previous discussions
the large imaginary parts of these scattering lengths have simply been ignored.
For example, Koike and Hayashigaki \cite{20} deduce a real $\rho N$
scattering length $a_{\rho N} \simeq 0.45 \, \fm$, so that the relation
${m_\rho^*}^2-m_\rho^2= -4\pi \rho (1 + m_\rho/M_N) a_{\rho N}$ implies a
downward mass shift of about $-45 \, \MeV$ at $\rho=\rho_0$. Our calculated
effective scattering length (\ref{4.12}) is qualitatively different.  We find a
large imaginary part $\Im \, a_{\rho N} \simeq 1\, \fm$ which reflects the
strong influence of the $\rho N \to \pi N, \, \pi \Delta$ inelastic channels.
The real part $\Re \, a_{\rho N}$, on the other hand, is about one order of
magnitude smaller than $\Im \, a_{\rho N}$. It may have different signs
depending on details of cutoff procedures, but this is almost irrelevant in
view of the large magnitude of the imaginary part.

Hatsuda, Lee  and Shiomi \cite{17} have pointed out that the low-density limit
$\delta m_V^2=-\rho \, T_{V N} $ is expected to break down already at a fraction of
nuclear matter density $\rho_0$. This is indeed the case, given that the
magnitude of $T_{VN}$ is large for $V=\rho, \, \omega$. It is necessary to use
the full vector meson propagator (\ref{3.30}), with the self-energy
$\Pi_V=-\rho \, T_{VN}$ iterated to all orders, and this has in fact been done
when deriving the spectral function (\ref{3.33}). Additional higher order
corrections in $\Pi_V$ as calculated e.g. in ref.\cite{35}, do not lead to
substantial changes for $\rho \lsim \rho_0$.

\section{\label{6} Summary and concluding remarks}

\noindent
(1) Guided by the chiral meson-baryon effective Lagrangian with inclusion of
    vector mesons and electromagnetic interactions, we have calculated the
    spectral distributions of current-current correlation functions in each of
    the $\rho$-, $\omega$-, and $\phi$-meson channels, both in the vacuum and
    in baryonic matter. We observe perfect consistency with QCD sum rules in
    all cases.\\
\\
\noindent
(2) Our results do not suggest a universal scaling law for in-medium vector
    meson masses. We find almost no mass shifts for $\rho$ and $\phi$-mesons
    in matter, whereas the $\omega$ meson experiences a substantial
    (attractive) shift.\\
\\
\noindent    
(3) The $\rho N$ and $\omega N$ effective scattering amplitudes are
    governed by large inelasticities. For the $\rho$ meson channel the
    inelastic broadening moves a large amount of the spectral strength down to
    the continuum below the free $\rho$ meson resonance. This broadening is so
    strong that the $\rho$ meson is unlikely to survive as a quasiparticle at
    densities much above $\rho_0$, the density of normal nuclear matter. In
    contrast, the $\omega$ meson, with its small width in vacuum to start with,
    can still be identified as a quasiparticle in matter at $\rho=\rho_0$. The
    $\phi N$ interaction is altogether weak. Consequently, the $\phi$ meson is
    expected to have almost no mass shift. Although its width at $\rho=\rho_0$
    increases by about an order of magnitude over its free width, it still
    persists as a relatively narrow resonance.\\
\\
\noindent
(4) To the extent that the changes of vector meson spectra in matter are driven
    by an effective chiral Lagrangian, they represent the dynamics and symmetry
    breaking pattern characteristic of low energy QCD. These features enter at
    the same time in the operator product expansion part of the QCD sum
    rules. Whether they directly reflect a tendency toward chiral symmetry
    restoration at high density is less obvious.\\
\\
\noindent
(5) Our calculations indicate that the lifetime of the $\omega$ meson in
    matter at $\rho=\rho_0$ decreases to about $\tau_\omega \sim 1.5$ fm/c. The
    production of ``slow'' $\omega$ mesons in a nuclear environment can
    therefore make in-medium changes of the $\omega$ meson spectrum
    visible. Such processes can be explored with the upcoming HADES lepton pair
    spectrometer at GSI. A reaction of particular interest is $\pi^- p \to
    \omega n$ in heavy nuclei with subsequent decay $\omega \to e^+ e^-$ and a
    kinematic cut focusing on low $\omega$ meson momenta \cite{38,39}.\\
\\
\noindent
(6) For the lifetime of the $\phi$ meson we still expect
    $\tau_\phi \sim 10$ fm/c at $\rho=\rho_0$ which exceeds nuclear
    dimensions. We mention that once the effective in-medium $K^+$ and $K^-$
    masses are introduced according to ref.\cite{40}, a slight downward shift
    of the $\phi$ meson is expected (by 1-2 \% at $\rho=\rho_0$), and its
    lifetime decreases to about 6 fm/c.\\
\\
\\
Acknowledgments:\\
We are grateful to Gerry Brown, Bengt Friman, Mannque Rho and Madeleine Soyeur for stimulating discussions.

\newpage
 
\underline{\Large Figure Captions:}\\

Figure 1: The ratio $\sigma(e^+e^- \to {\rm hadrons})/\sigma(e^+e^- \to \mu^+
\mu^-)$ as a function of the total c.m. energy. The isovector, isoscalar and
the $\phi$ meson channels, corresponding to the currents (3-5), are shown
separately in Figs. 1a, b and c, respectively. The data are taken from
refs.[30-35]. In the $\phi$ channel the high energy data sample summarizes
$K\bar{K}+n\,\pi$ final states. The solid lines are the results of the VMD
model. The dashed lines represent the perturbative QCD limit.\\

Figure 2: Diagrammatic representation of the full photon propagator (wavy line)
and of the full vector meson (V) propagator.\\

Figure 3: Comparison of the Borel transformed right side (operator product
expansion, OPE) and left side (dispersion relation, DR) of the QCD sum rules
for the isovector (I=1), isoscalar (I=0) and $\phi$ meson channels, as a
function of the Borel mass $\cal M$ (see eq.(\ref{2.20})). The solid line is
the result using our spectral density which describes the experimental data.
The short dashed line is the simplified ansatz using a delta function for
each vector meson resonance. \\

Figure 4: Feynman rules for the vertex functions derived from the effective
Lagrangian of the coupled system of vector mesons, pseudoscalar mesons and
baryons. Examples are shown here for interactions involving $\rho$ mesons and
pions (including the $\pi \rho \omega$ coupling). Solid lines: nucleons; double
solid lines: $\Delta$-isobars; dashed lines: pions; vector mesons are drawn as
``zig zag'' lines. The generalization to flavour SU(3) is straightforward. Here
$S$ and $T$ are the $\Delta \to N$ spin and isospin transition operators and
$\vec{\Theta}$ is the isospin operator for the $\Delta$ (see ref.\cite{41}).\\

Figure 5: Processes contributing to the $\rho N$ scattering amplitude. The
dashed lines represents pions, the wavy line describes the $\rho$ meson. The
incoming baryon (solid line) is always a nucleon while the intermediate baryon
can be either a nucleon or a delta.\\

Figure 6: Two-loop contribution to the $\rho N$ scattering amplitude including
$\pi \pi $ interactions.\\

Figure 7: Imaginary parts of the $V N$ scattering amplitudes in the isovector
(a), isoscalar (b) and $\phi$ meson (c) channels. Solid lines: total
results. Dashed lines show decomposition into selected subchannels.\\

Figure 8: Real and imaginary parts of $V N$ scattering amplitudes in different
channels. Solid lines: imaginary parts used as input for dispersion relations
to generate the real parts shown as long-dashed lines. Fig. 8a also shows
separate contribution from $\rho N \to \pi N $ (with $g_A=1$) and $\rho N \to
\pi N + \pi \Delta$ (with full axial formfactor).\\

Figure 9: Spectra of the current-current correlation functions in different
channels, shown at zero (dashed lines), half (long dashed) and normal nuclear
density (solid lines). \\

Figure 10: Real part of in-medium vector meson propagator. The lines
correspond to zero (dashed), half (long dashed) and normal nuclear
density (solid). \\

Figure 11: The Borel transformed ``left side'' (l.s.) (from dispersion relation,
dot dashed) and ``right side'' (r.s.) (from OPE, solid line) of  the QCD sum rule (see
eq.(\ref{5.5})) at normal nuclear
density. For comparison we also show the vacuum results of Fig. 3.\\

\newpage
\begin{figure}[h]
\unitlength1mm
\begin{picture}(100,225)
\put(0,0){\framebox(100,225){}}
\put(0,75){\line(1,0){100}}
\put(0,150){\line(1,0){100}}
\put(5,0){\makebox{\epsfig{file=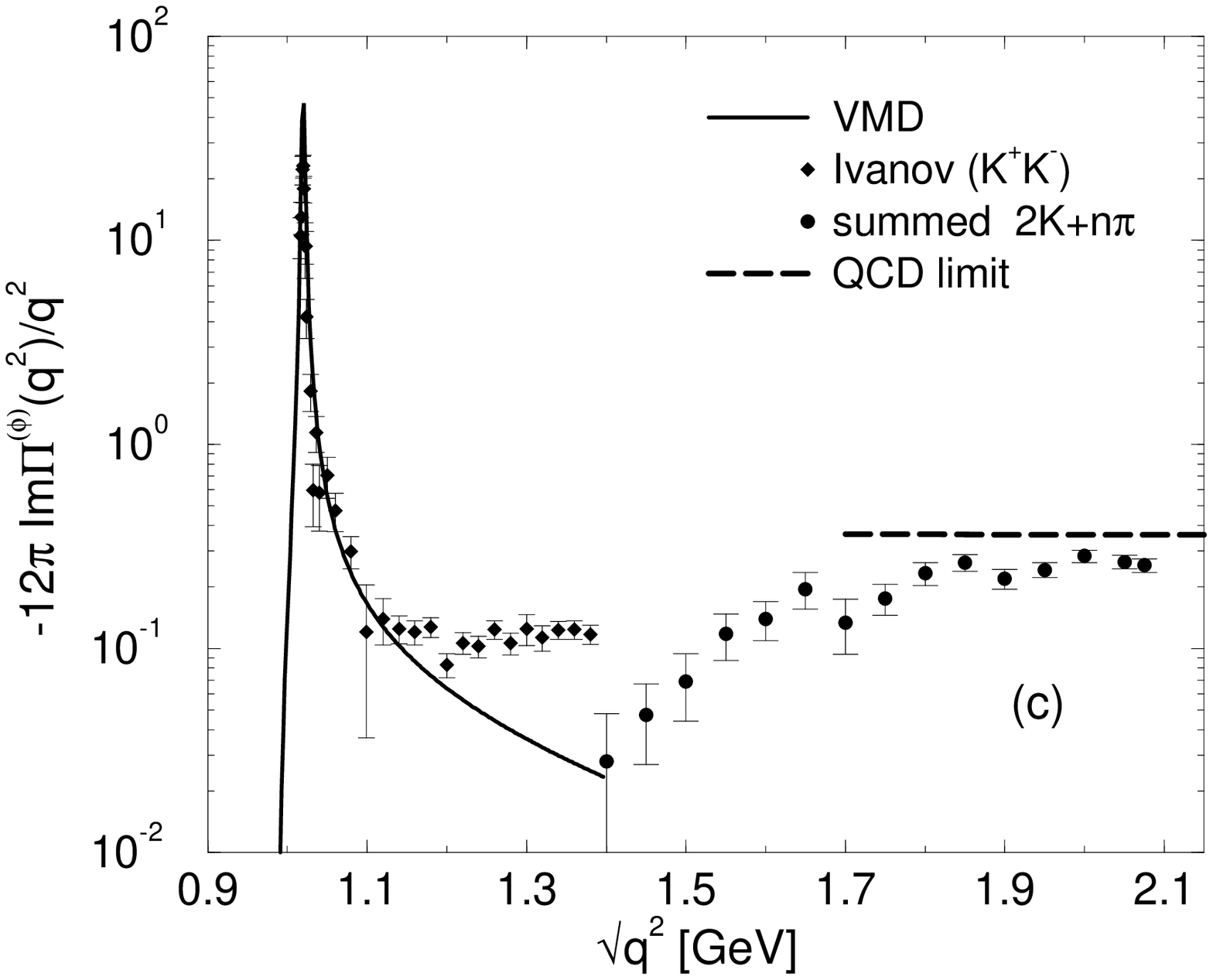,width=95mm}}}
\put(5,75){\makebox{\epsfig{file=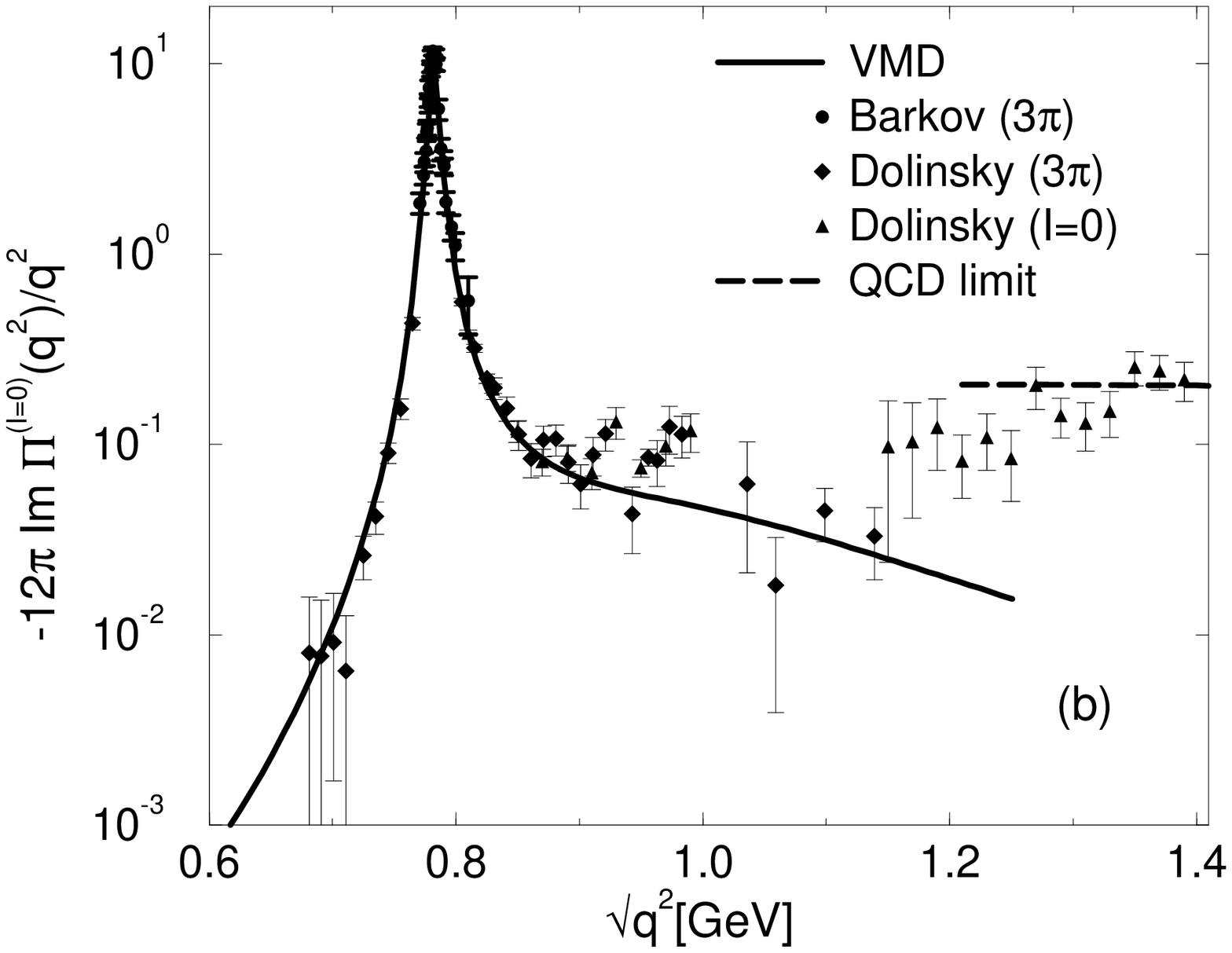,width=95mm}}}
\put(5,150){\makebox{\epsfig{file=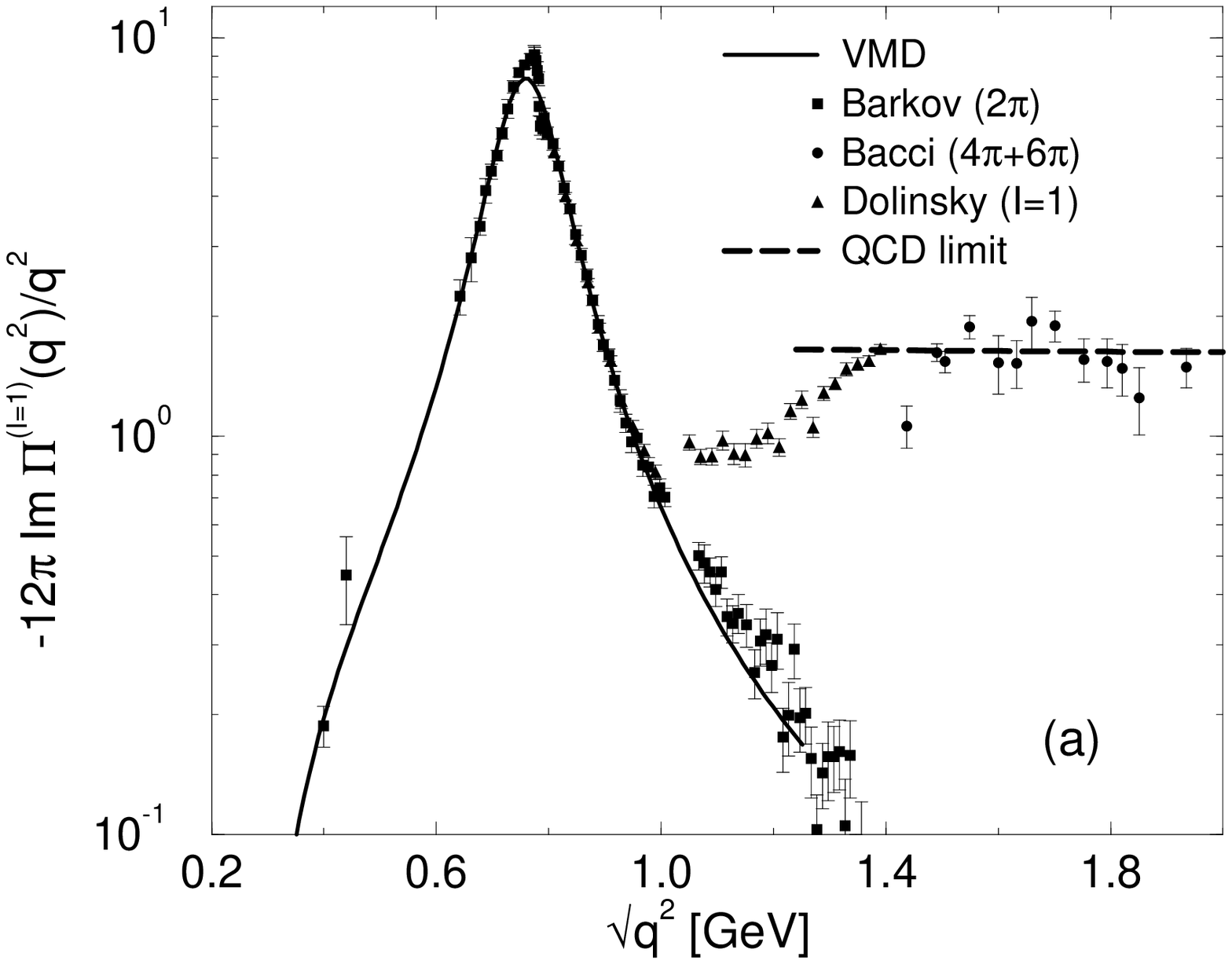,width=95mm}}}
\end{picture}
Fig.1
\end{figure}

\begin{figure}[h] 
\unitlength1mm
\begin{picture}(100,100)
\put(5,0){\makebox{\epsfig{file=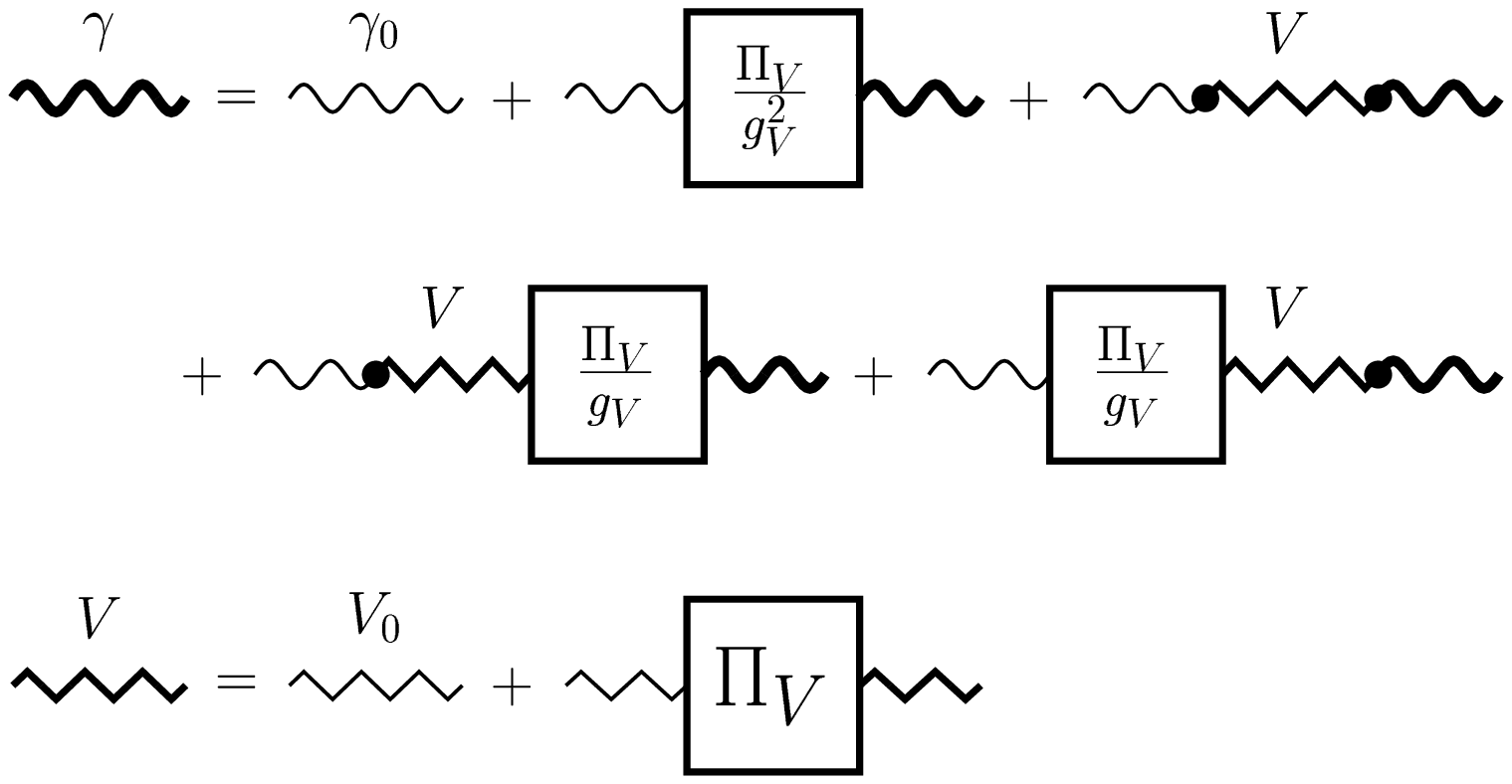,width=95mm}}}
\end{picture}
Fig.2
\end{figure}

\begin{figure}[h]
\unitlength1mm
\begin{picture}(100,225)
\put(0,0){\framebox(100,225){}}
\put(0,75){\line(1,0){100}}
\put(0,150){\line(1,0){100}}
\put(5,0){\makebox{\epsfig{file=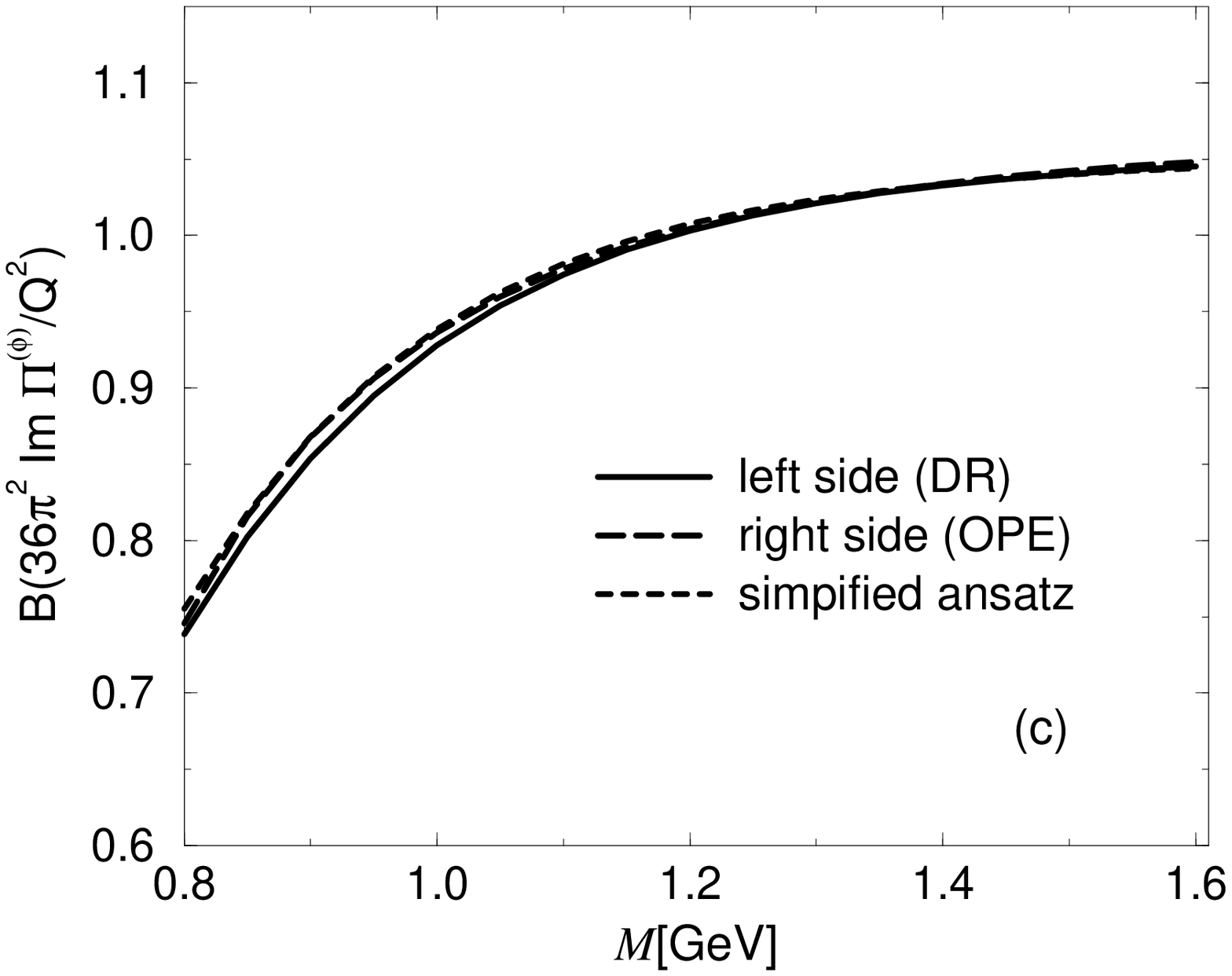,width=95mm}}}
\put(5,75){\makebox{\epsfig{file=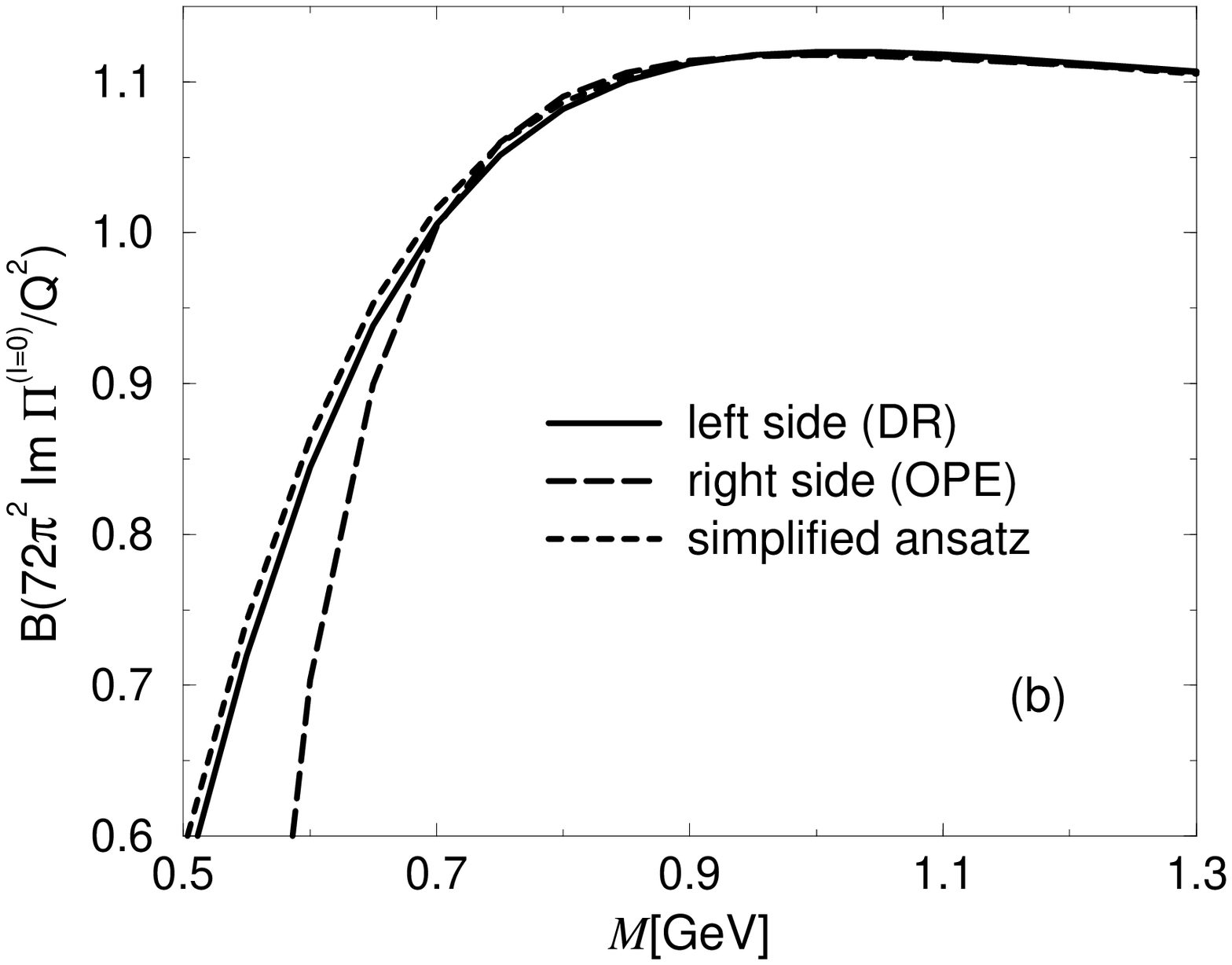,width=95mm}}}
\put(5,150){\makebox{\epsfig{file=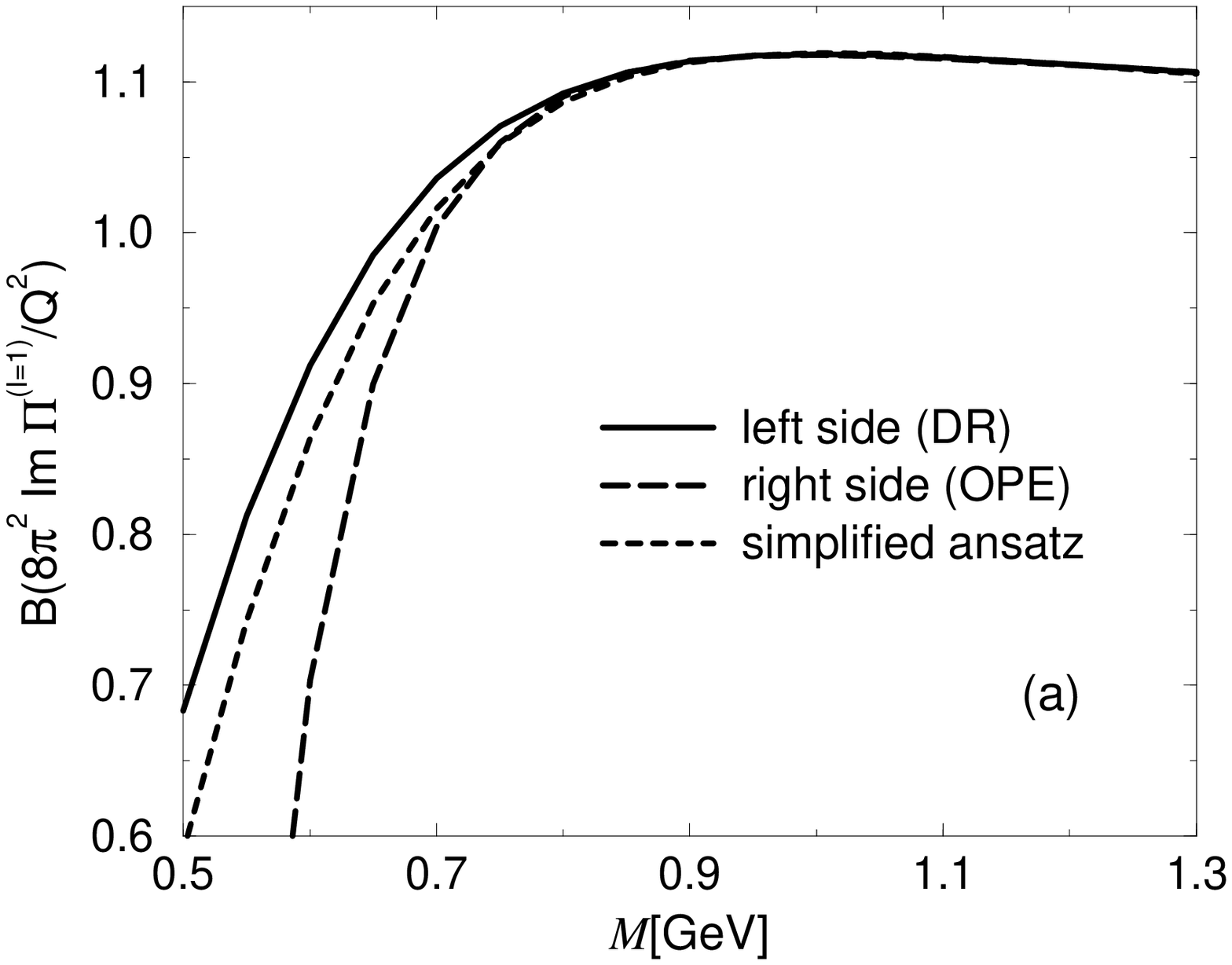,width=95mm}}}
\end{picture}
Fig.3
\end{figure}

\begin{figure}[h] 
\unitlength1mm
\begin{picture}(100,200)
\put(5,100){\makebox{\epsfig{file=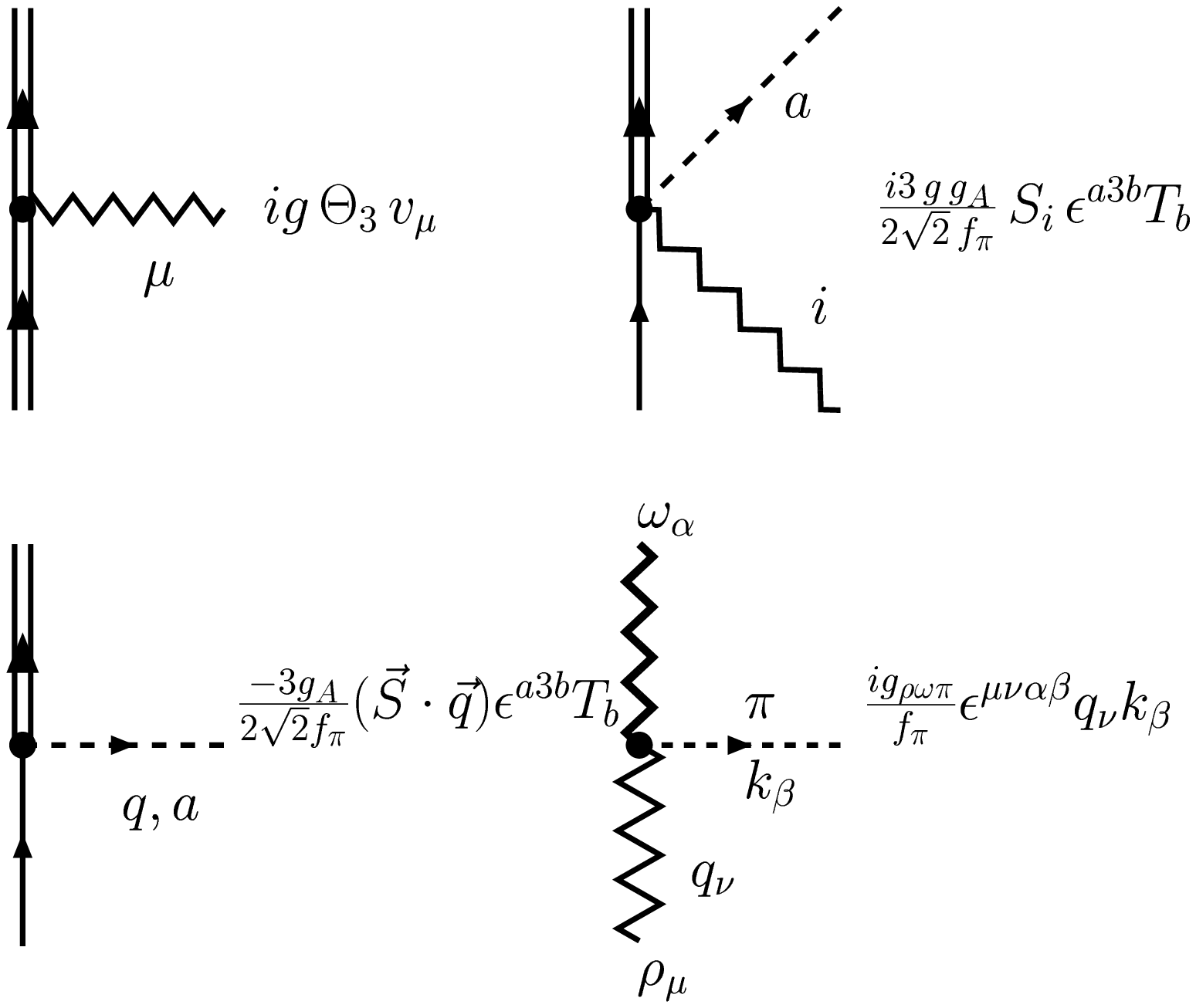,width=80mm}}}
\put(5,0){\makebox{\epsfig{file=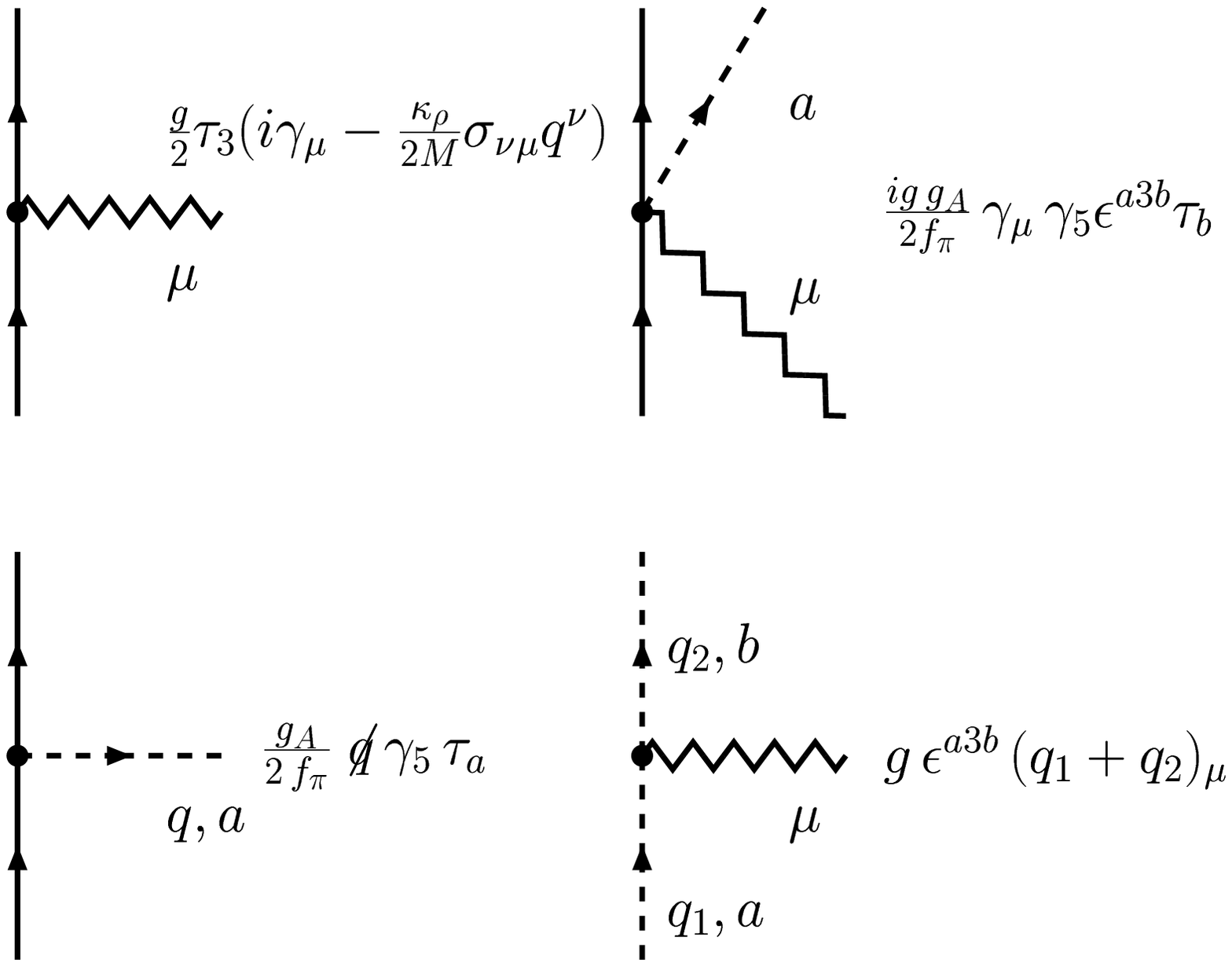,width=80mm}}}
\end{picture}
Fig.4
\end{figure}

\begin{figure}[h]
\vspace*{-10cm} 
\unitlength1mm
\begin{picture}(100,200)
\put(5,0){\makebox{\epsfig{file=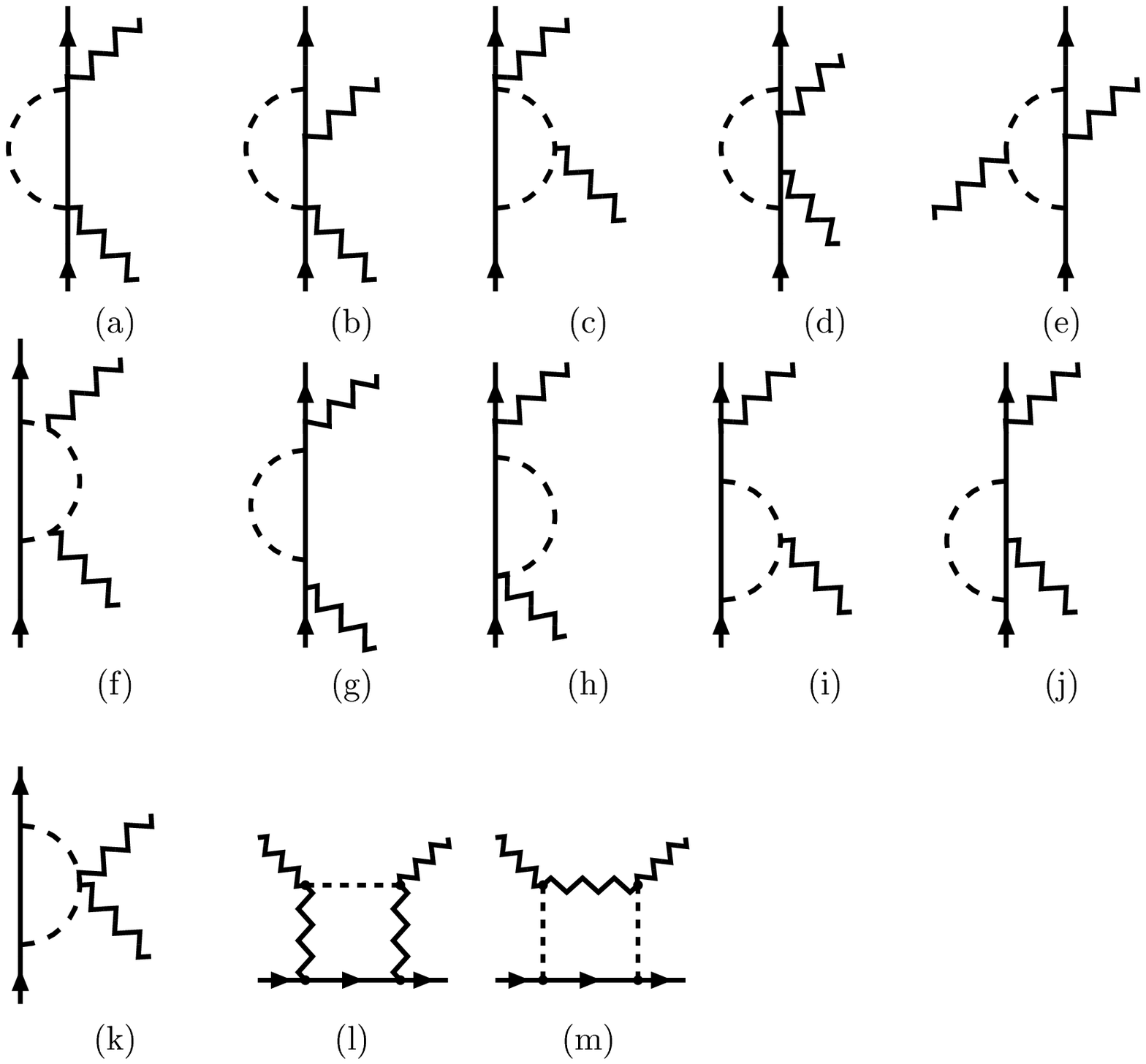,width=80mm}}}
\end{picture}
Fig.5
\end{figure}

\begin{figure}[h] 
\unitlength1mm
\begin{picture}(50,50)
\put(5,0){\makebox{\epsfig{file=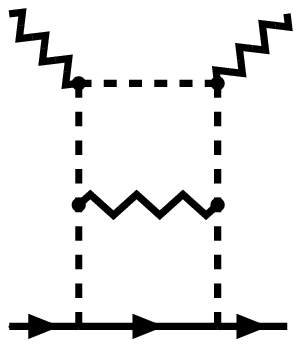,width=20mm}}}
\end{picture}
Fig.6
\end{figure}

\begin{figure}[h]
\unitlength1mm
\begin{picture}(100,225)
\put(0,0){\framebox(100,225){}}
\put(0,75){\line(1,0){100}}
\put(0,150){\line(1,0){100}}
\put(5,0){\makebox{\epsfig{file=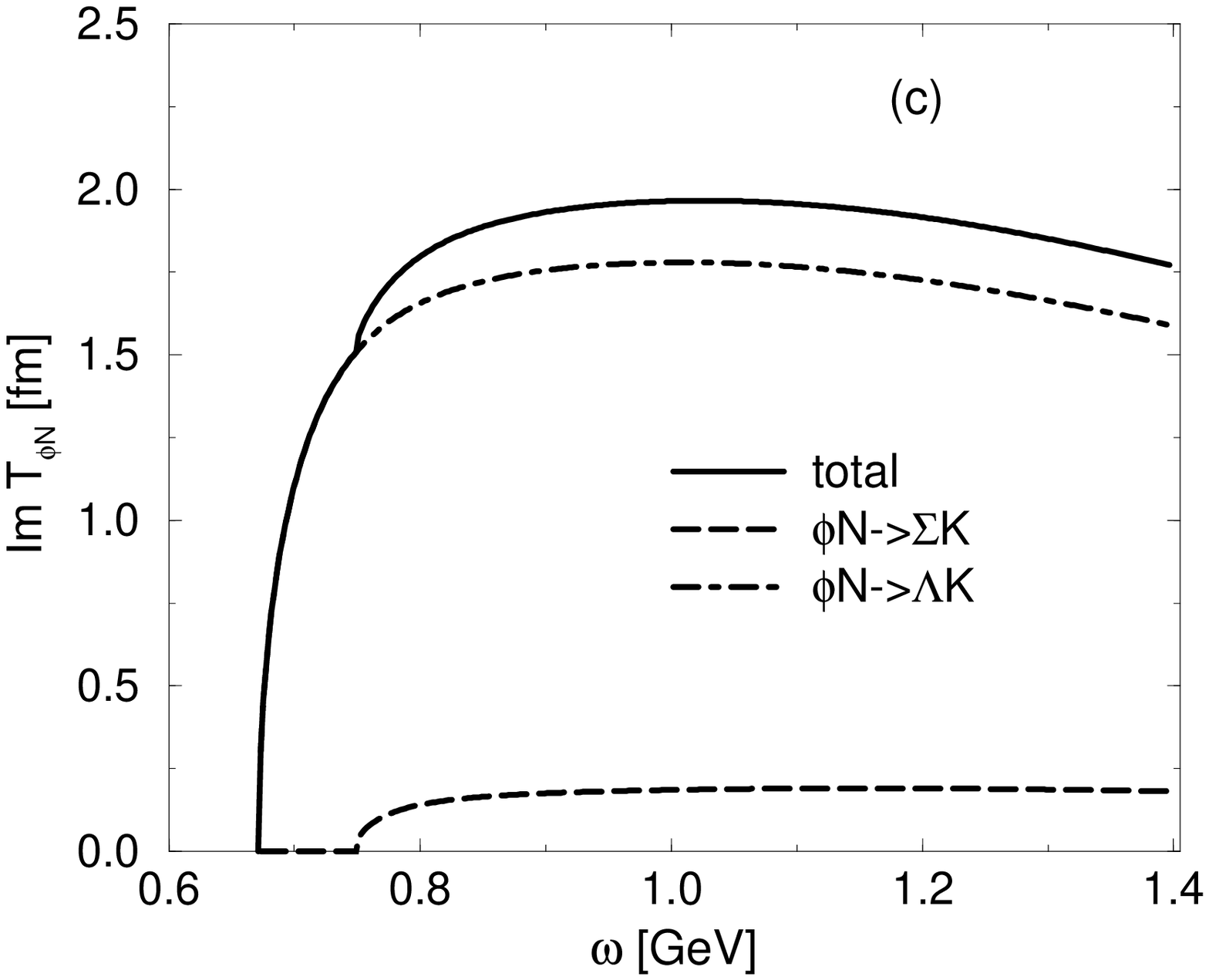,width=95mm}}}
\put(5,75){\makebox{\epsfig{file=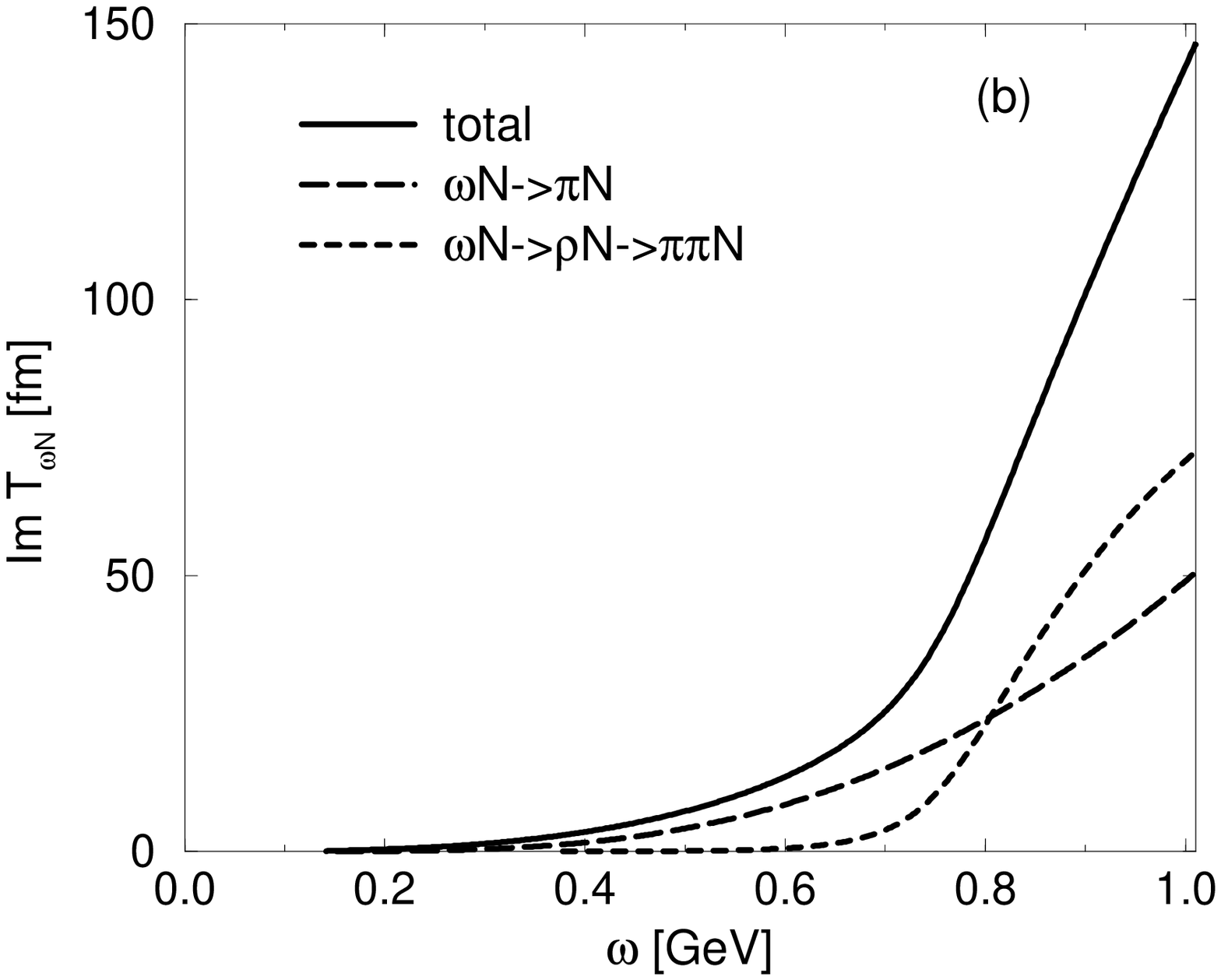,width=95mm}}}
\put(5,150){\makebox{\epsfig{file=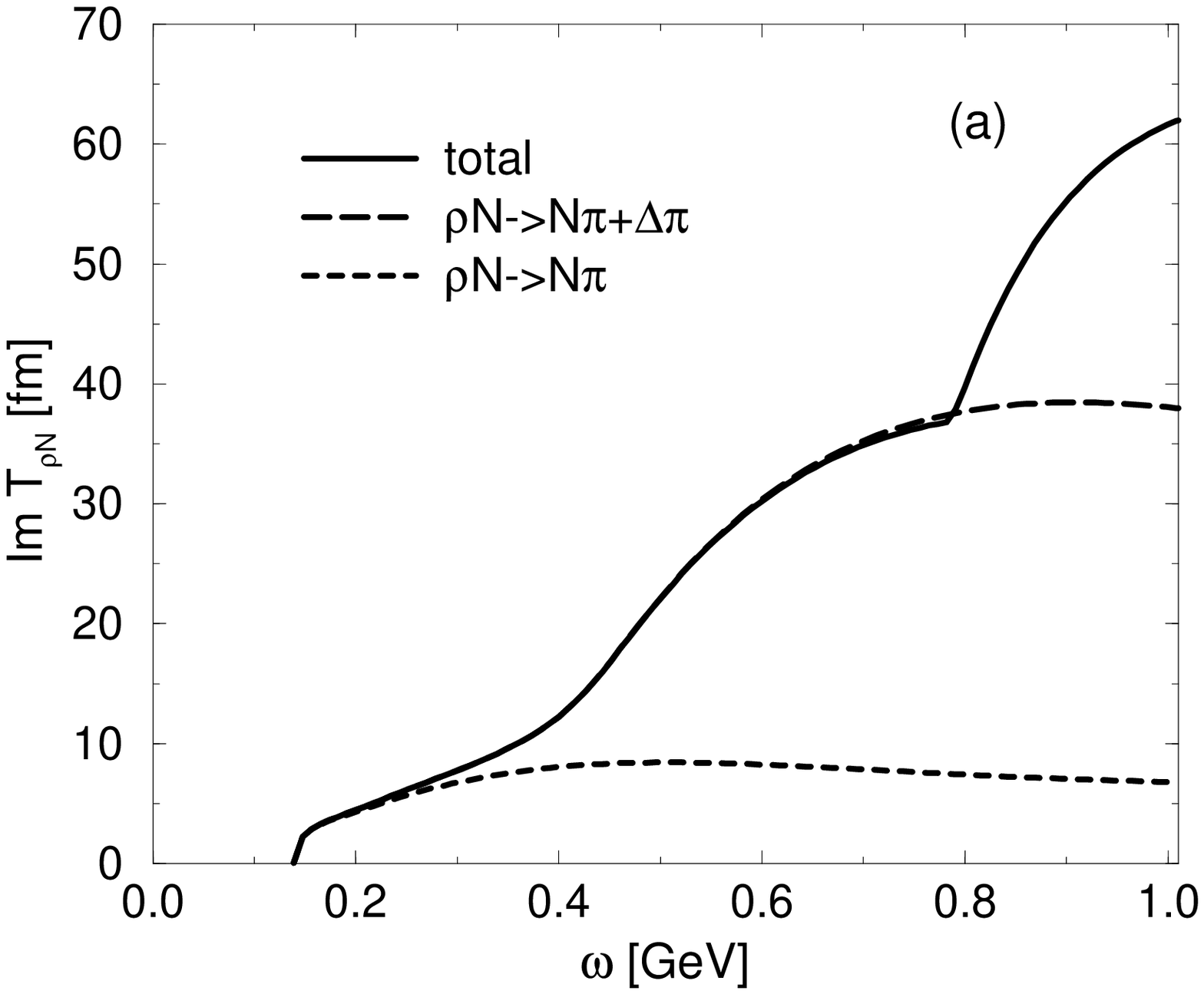,width=95mm}}}
\end{picture}
Fig.7
\end{figure}

\begin{figure}[h]
\unitlength1mm
\begin{picture}(100,225)
\put(0,0){\framebox(100,225){}}
\put(0,75){\line(1,0){100}}
\put(0,150){\line(1,0){100}}
\put(5,0){\makebox{\epsfig{file=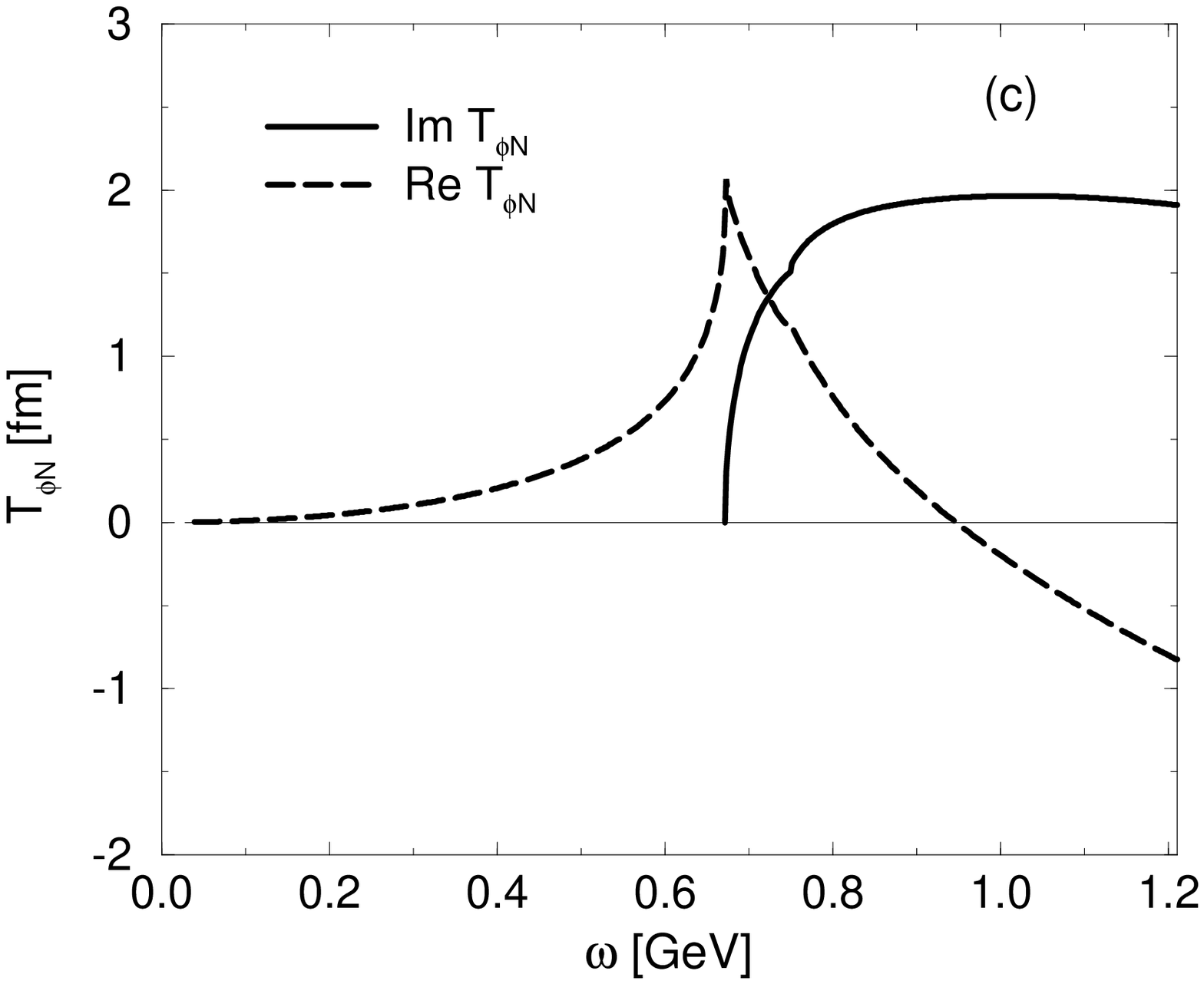,width=95mm}}}
\put(5,75){\makebox{\epsfig{file=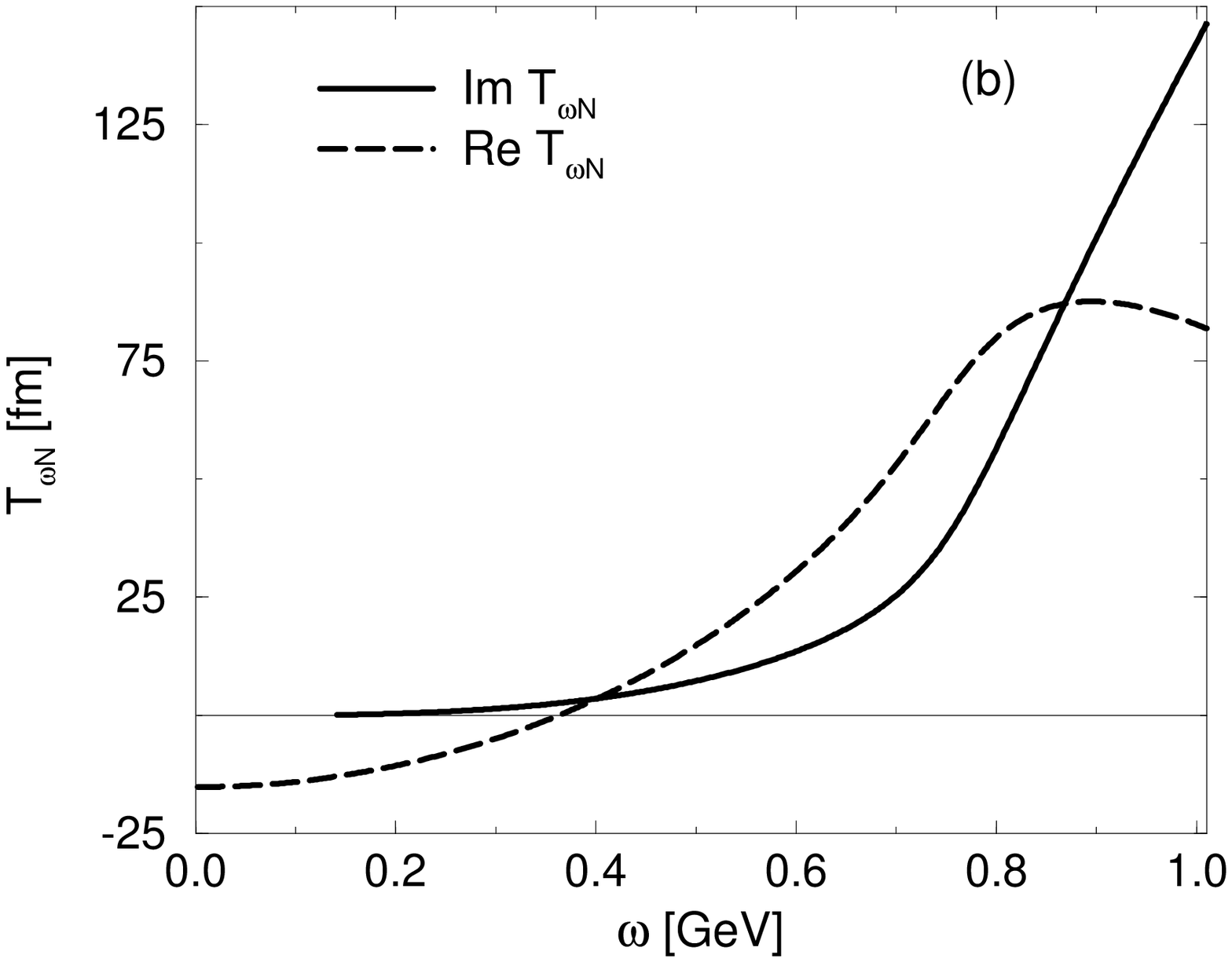,width=95mm}}}
\put(5,150){\makebox{\epsfig{file=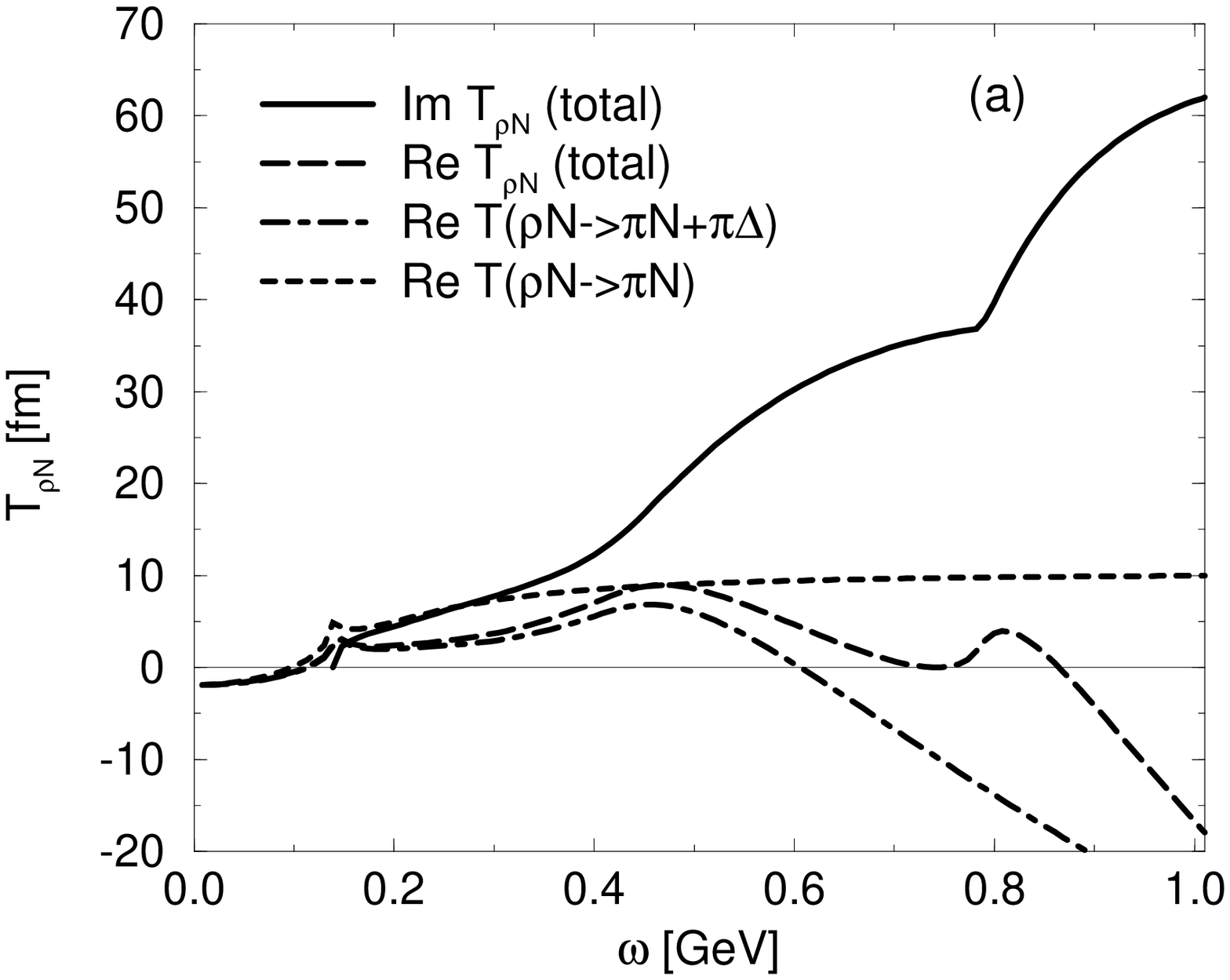,width=95mm}}}
\end{picture}
Fig.8
\end{figure}

\begin{figure}[h]
\unitlength1mm
\begin{picture}(100,225)
\put(0,0){\framebox(100,225){}}
\put(0,75){\line(1,0){100}}
\put(0,150){\line(1,0){100}}
\put(5,0){\makebox{\epsfig{file=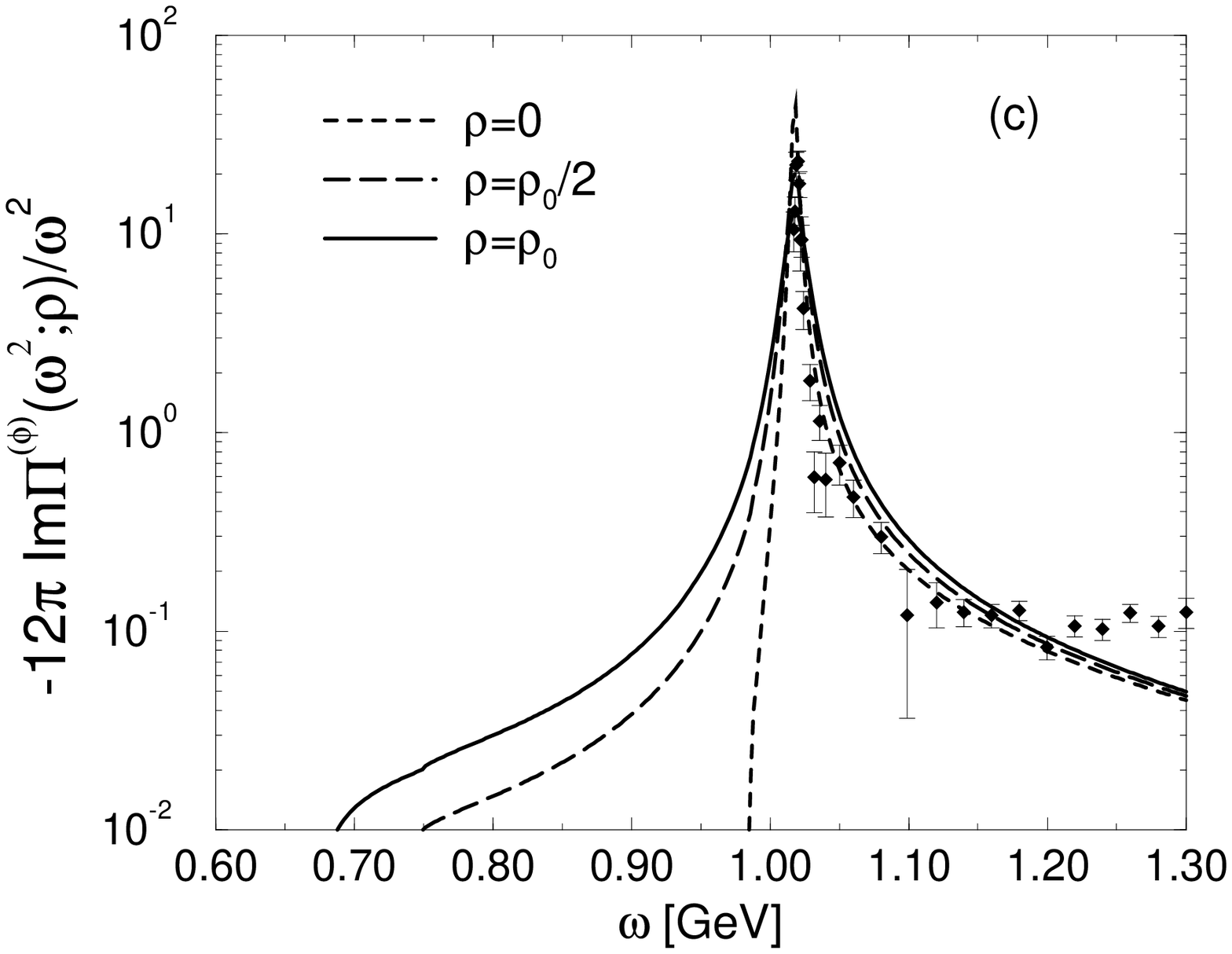,width=95mm}}}
\put(5,75){\makebox{\epsfig{file=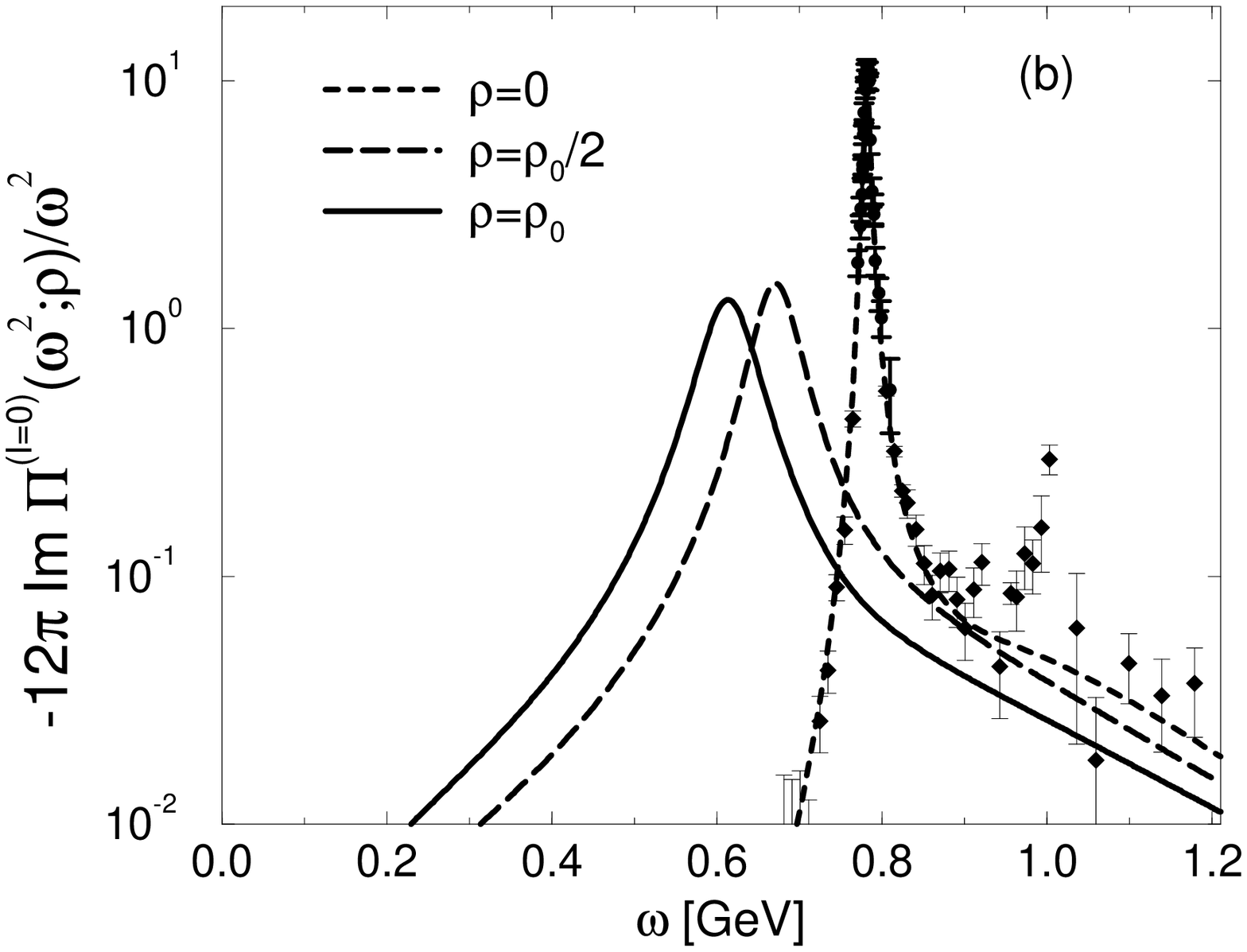,width=95mm}}}
\put(5,150){\makebox{\epsfig{file=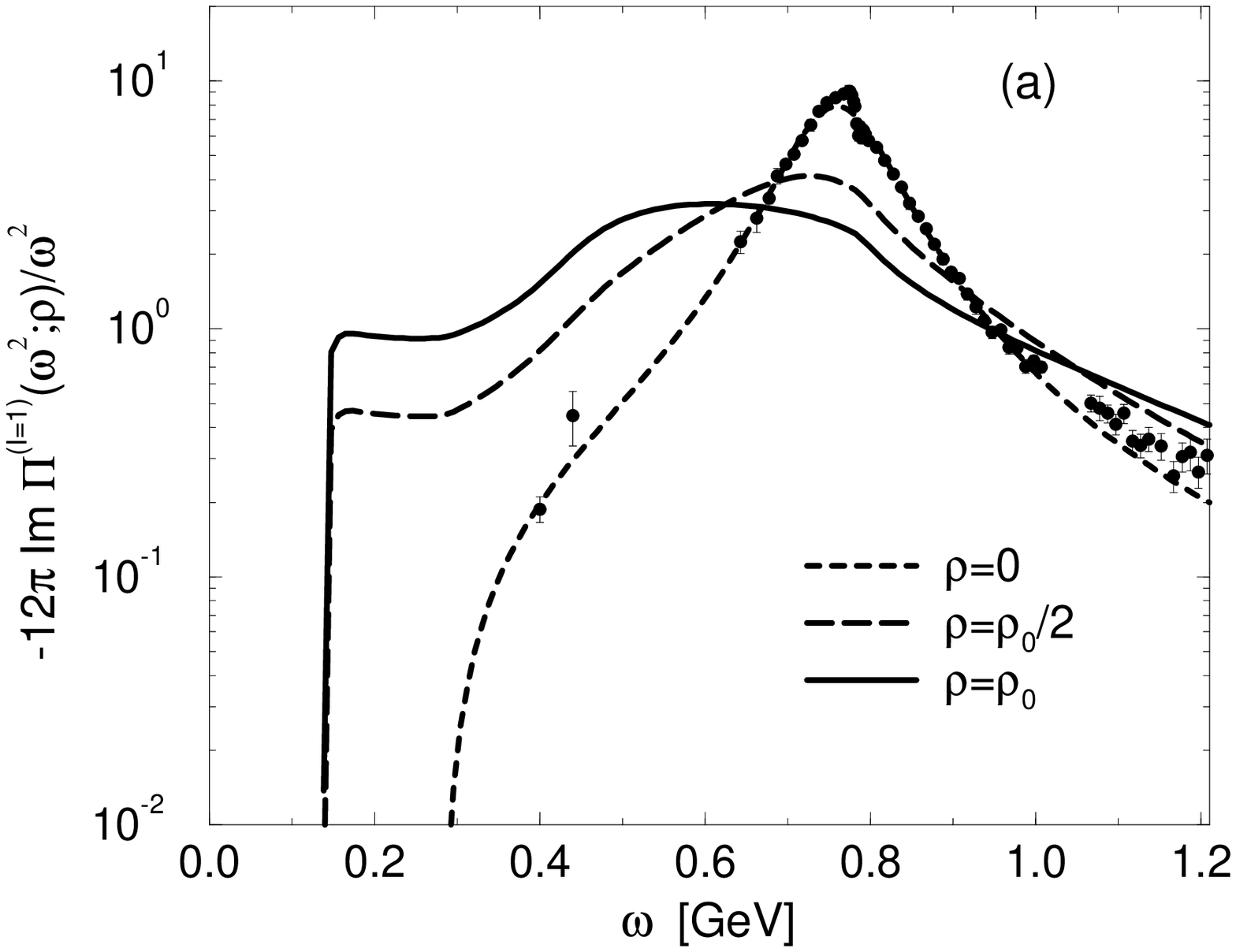,width=95mm}}}
\end{picture}
Fig.9
\end{figure}

\begin{figure}[h]
\unitlength1mm
\begin{picture}(100,225)
\put(0,0){\framebox(100,225){}}
\put(0,75){\line(1,0){100}}
\put(0,150){\line(1,0){100}}
\put(5,0){\makebox{\epsfig{file=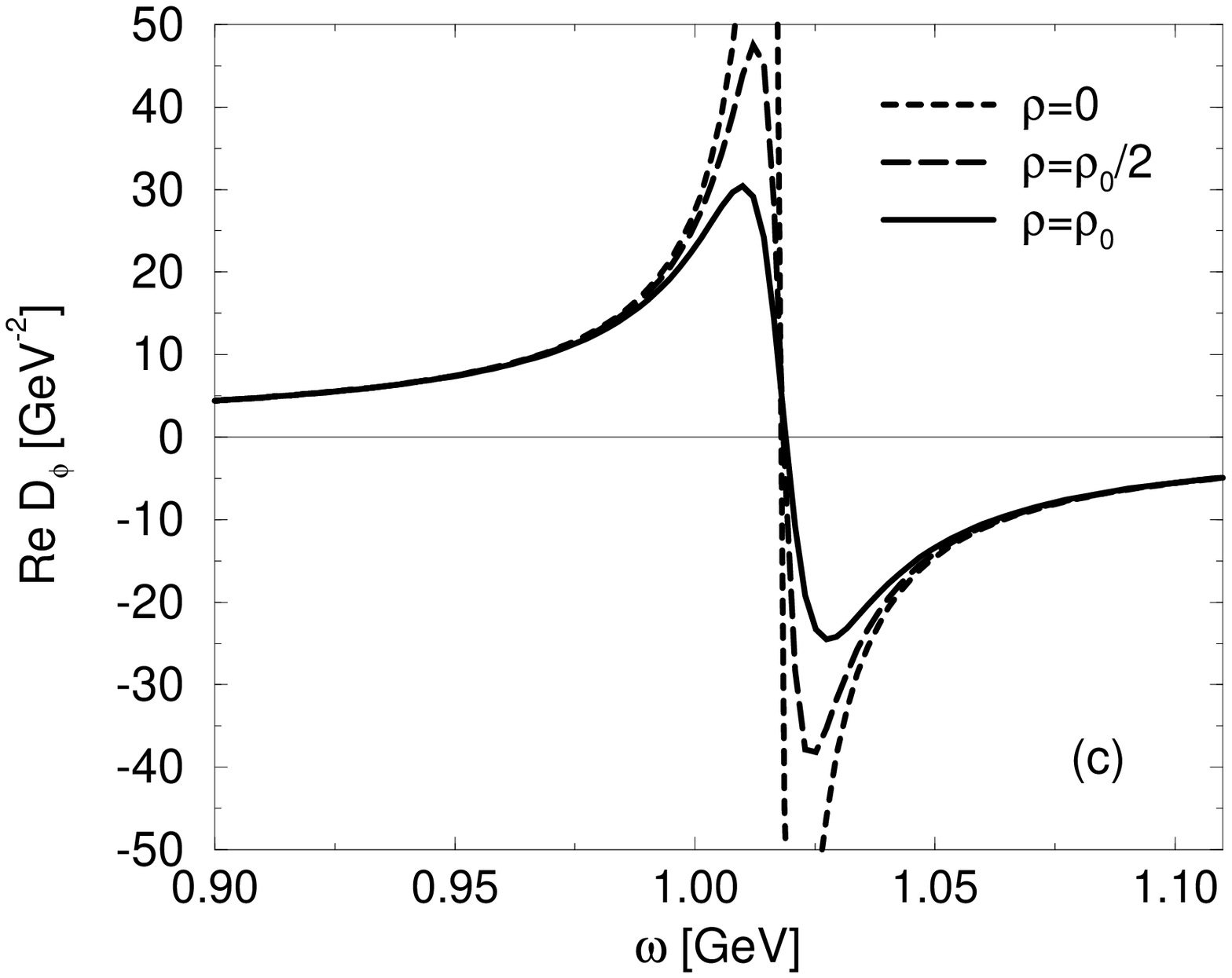,width=95mm}}}
\put(5,75){\makebox{\epsfig{file=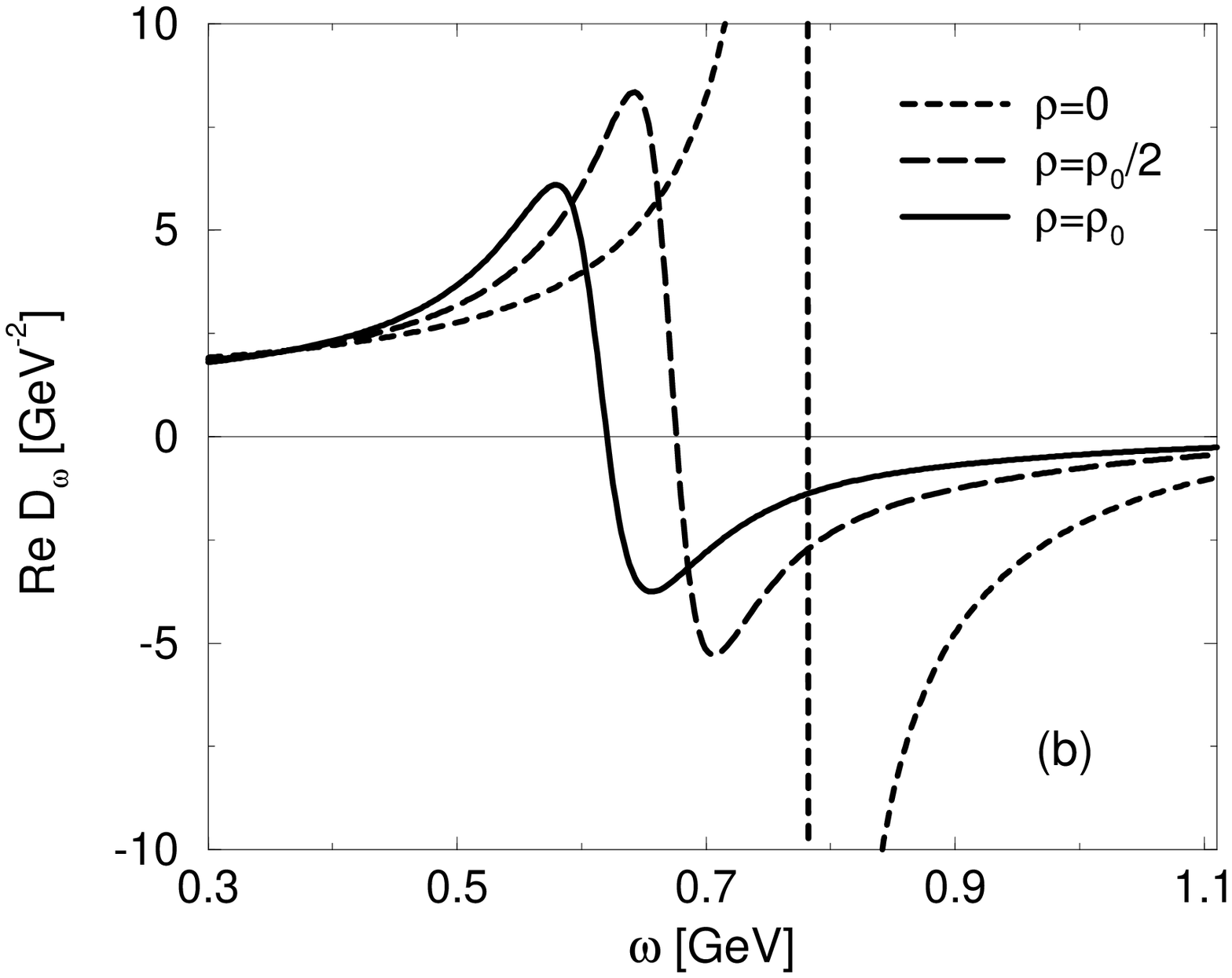,width=95mm}}}
\put(5,150){\makebox{\epsfig{file=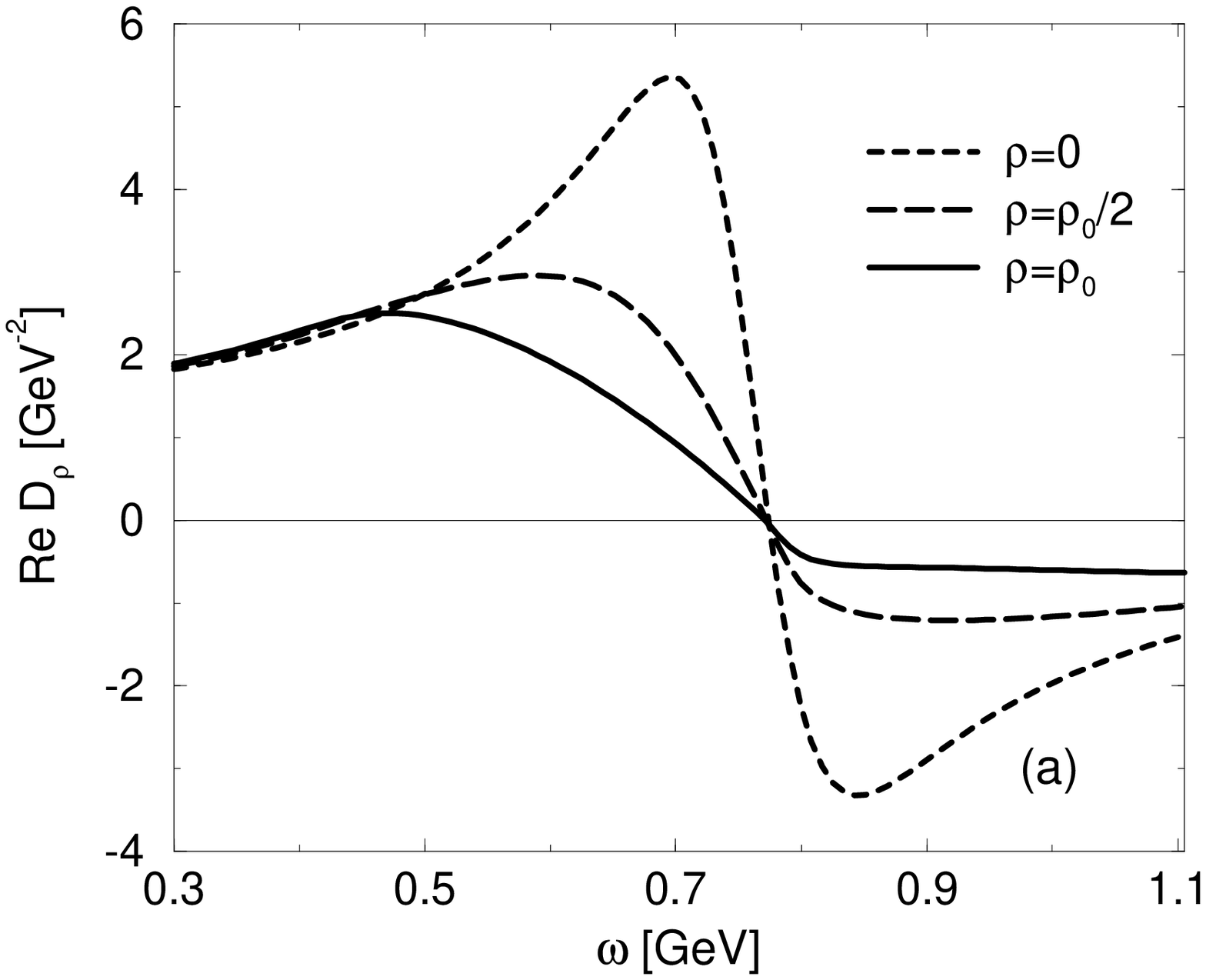,width=95mm}}}
\end{picture}
Fig.10
\end{figure}

\begin{figure}[h]
\unitlength1mm
\begin{picture}(100,225)
\put(0,0){\framebox(100,225){}}
\put(0,75){\line(1,0){100}}
\put(0,150){\line(1,0){100}}
\put(5,0){\makebox{\epsfig{file=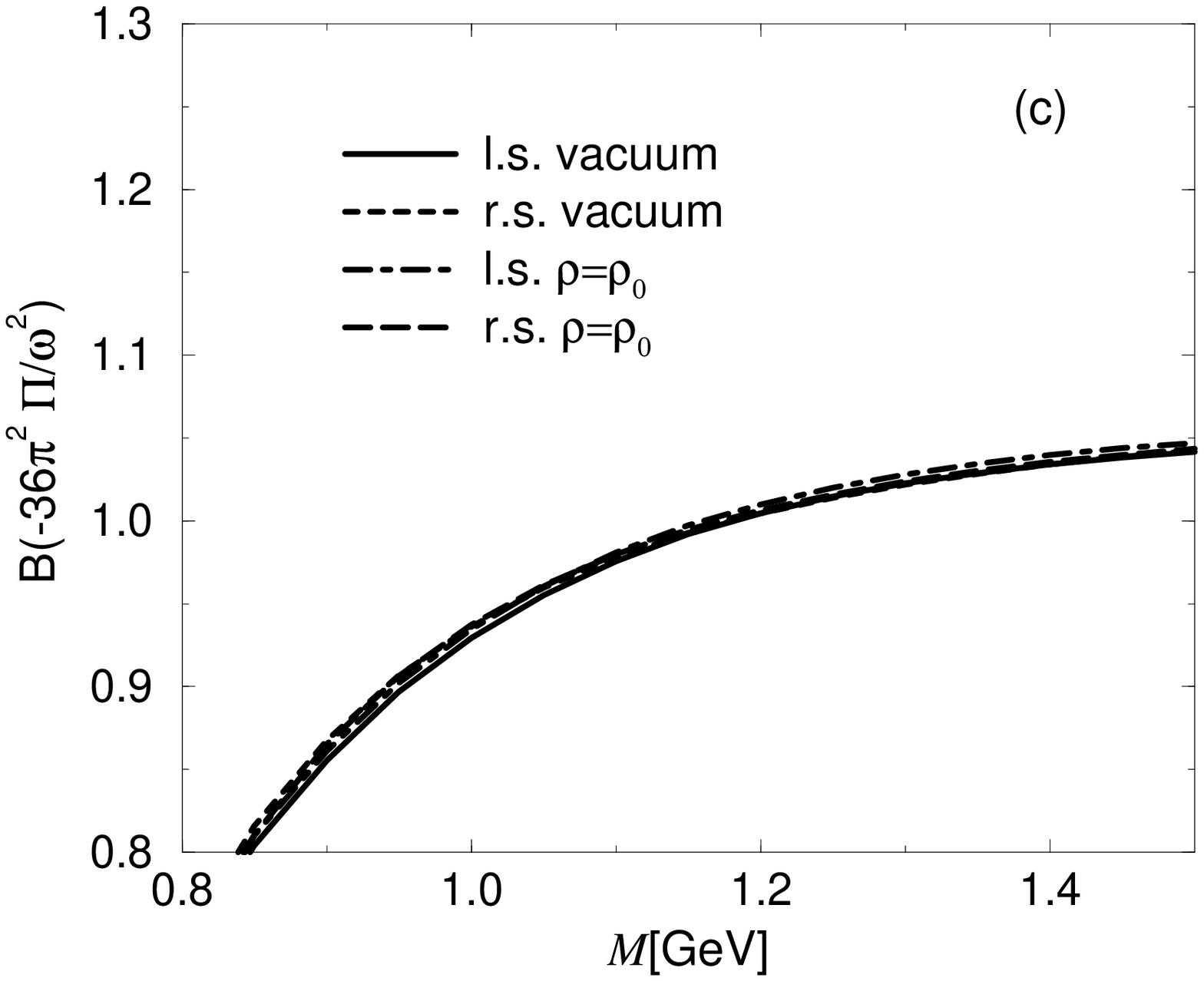,width=95mm}}}
\put(5,75){\makebox{\epsfig{file=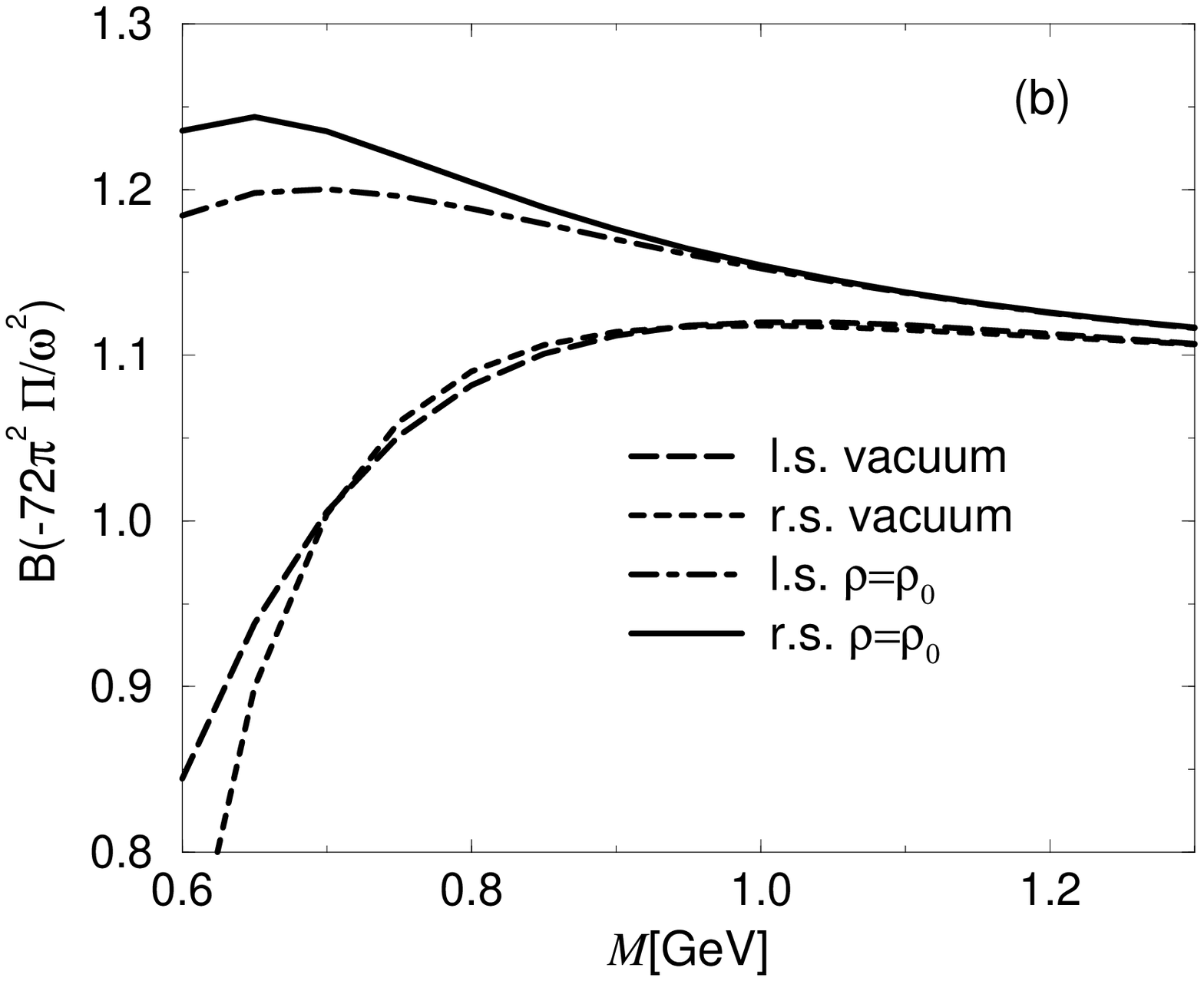,width=95mm}}}
\put(5,150){\makebox{\epsfig{file=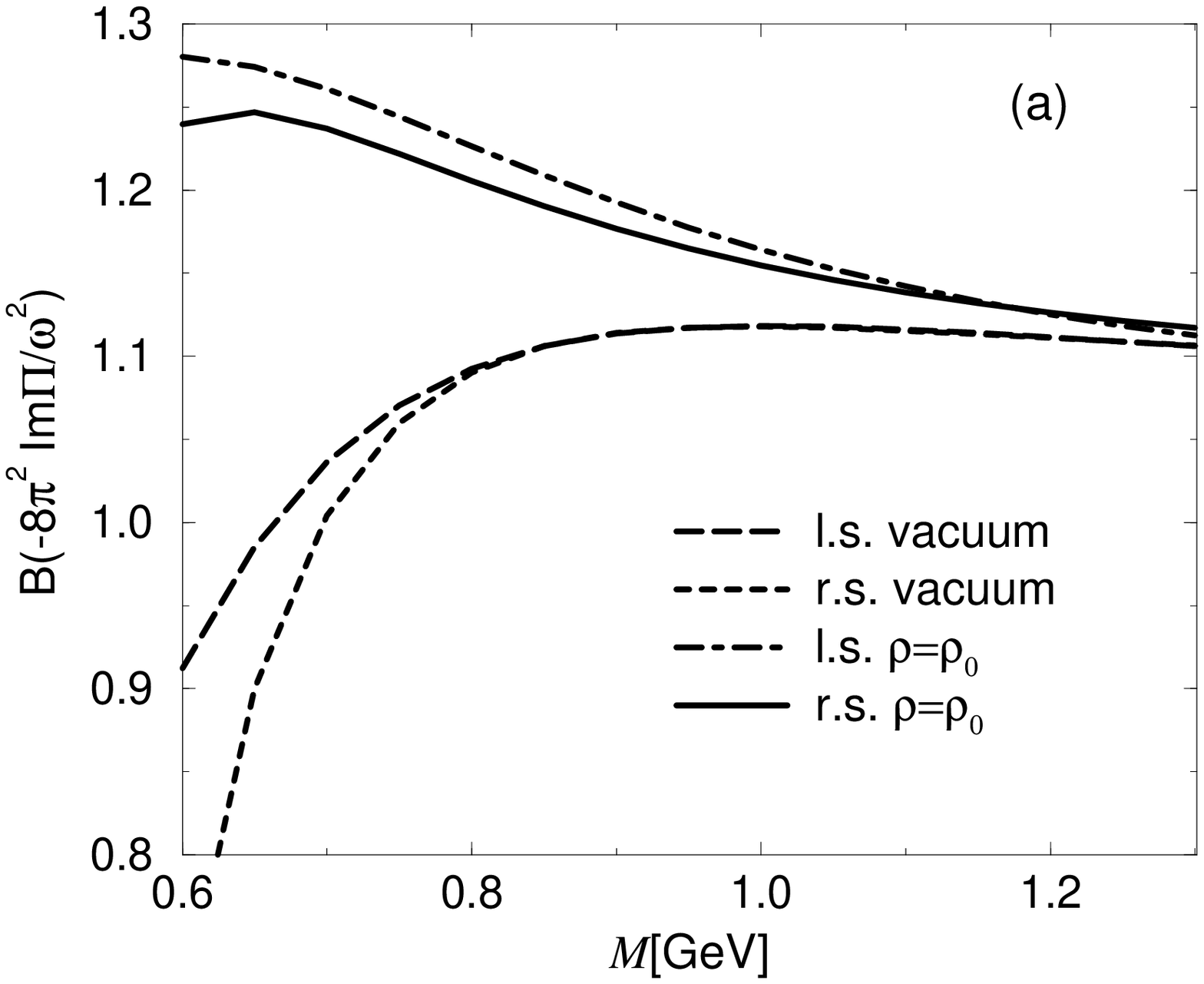,width=95mm}}}
\end{picture}
Fig.11
\end{figure}

\end{document}